\documentclass[prl,twocolumn,notitlepage,superscriptaddress,longbibliography]{revtex4-1}
\usepackage{amsmath}
\usepackage{amssymb}

\usepackage[unicode=true,colorlinks=true,citecolor=blue,urlcolor=blue]{hyperref}

\usepackage{bm}
\usepackage{color}
\usepackage{epsfig}

\usepackage[normalem]{ulem}

\renewcommand {\Im}{\mathop\mathrm{Im}\nolimits}
\renewcommand {\Re}{\mathop\mathrm{Re}\nolimits}
\newcommand {\Tr}{\mathop\mathrm{Tr}\nolimits}
\renewcommand {\i}{{\rm i}}
\renewcommand {\phi}{{\varphi}}
\newcommand {\rmi}{{\rm i}}
\newcommand {\rmd}{{\rm d}}
\newcommand {\sign}{\mathop{\mathrm{sign}}\nolimits}
\newcommand {\e}{{\rm e}}
\newcommand {\eps}{\varepsilon}
\newcommand{\done}{\marginpar{\checkmark}}
\newcommand {\rot}{\mathop\mathrm{rot}\nolimits}
\begin{document}
	\title{Inelastic scattering of photon pairs in qubit arrays with subradiant states
		%Inelastic light scattering in qubit arrays  enhanced by multiexcitation subradiant  states
		%Role of multi-excited subradiant state and twilight state on inelastic scattering
		%Subradiant and twilight states of  multiply photons  interacting with  qubits in a waveguide//
	}
	%(order of authors and affiliations is tentative)
	\author{Yongguan Ke}
	\affiliation{Laboratory of Quantum Engineering and Quantum Metrology, School of Physics and Astronomy,
		Sun Yat-Sen University (Zhuhai Campus), Zhuhai 519082, China}
	\affiliation{Nonlinear Physics Centre, Research School of Physics, Australian National University, Canberra ACT 2601, Australia}
	\author{Alexander V. Poshakinskiy}
	\affiliation{Ioffe Institute, St. Petersburg 194021, Russia}
	\author{Chaohong Lee}
	\email{lichaoh2@mail.sysu.edu.cn}
	\affiliation{Laboratory of Quantum Engineering and Quantum Metrology, School of Physics and Astronomy,
		Sun Yat-Sen University (Zhuhai Campus), Zhuhai 519082, China}
	\affiliation{State Key Laboratory of Optoelectronic Materials and Technologies, Sun Yat-Sen University (Guangzhou Campus), Guangzhou 510275, China}
	%\affiliation{Nonlinear Physics Centre, Canberra ACT 2601, Australia}
	
	\author{Yuri S. Kivshar}
	\affiliation{Nonlinear Physics Centre, Research School of Physics, Australian National University, Canberra ACT 2601, Australia}
	\affiliation{ITMO University, St. Petersburg 197101, Russia}
	\author{Alexander N. Poddubny}
	\email{poddubny@coherent.ioffe.ru}
	\affiliation{Nonlinear Physics Centre, Research School of Physics, Australian National University, Canberra ACT 2601, Australia}
	\affiliation{Ioffe Institute, St. Petersburg 194021, Russia}
	\affiliation{ITMO University, St. Petersburg 197101, Russia}
\begin{abstract}
	We  develop a rigorous  theoretical  approach for analyzing inelastic scattering of photon pairs in arrays of two-level qubits embedded into a waveguide. Our analysis reveals  strong enhancement of the scattering  when the energy of incoming photons resonates with the double-excited subradiant states. We identify the role of different  double-excited states in the scattering such as {\em superradiant}, {\em subradiant}, and  {\it twilight} states, being a product of single-excitation bright and subradiant states. Importantly,  the $N$-excitation subradiant states can be engineered only if the number of qubits exceeds $2N$. Both the subradiant and twilight states can generate long-lived photon-photon correlations, paving the way to a storage and processing of quantum information.
\end{abstract}
\date{\today}

\maketitle
%%%%%%%%%%%%%%%%%%%%%%%%%%%%%%%%%%%%%%%%%%
{\it Introduction}.
Nonlinear manipulation of light via its interaction with matter plays an essential role in optics and its applications~\cite{chang2014,Konstantin2016,Kruk2019}, including optical communications~\cite{kimble2008quantum} and sensing~\cite{Tittl2018}.
The light-matter interaction can be strongly modified  by collective coherent superradiance or subradiance, where the spontaneous emission
speeds up or slows down~\cite{Dicke1954,Roy2017,KimbleRMP2018,FriskKockum2019}.
Both  superradiance and subradiance have been realized in various systems~\cite{Ivchenko1994,Birkl1995,Devoe1996,chumakov1999,Hendrickson2008,Goldberg2009,vanLoo2013,Ivchenko2013,
	mlynek2014,Guerin2016,Jenkin2017,Limonov2017,Wolf2018,Weiss2018,Wang2019arxiv}, and they provide novel opportunities to explore the interplay between collective excitations in materials and nonlinear effects in scattering of light~\cite{Roy2017,KimbleRMP2018}.
Compared to superradiance, subradiance enables longer time for light-matter interaction, and giant nonlinear response~\cite{Carletti2018,ANP2018,Koshelev2019}. To the best of our knowledge, the enhancement of light-matter interaction
by subradiant modes has been explored mostly in classical optics.

\begin{figure}[!b]
	\centering\includegraphics[width=0.48\textwidth]{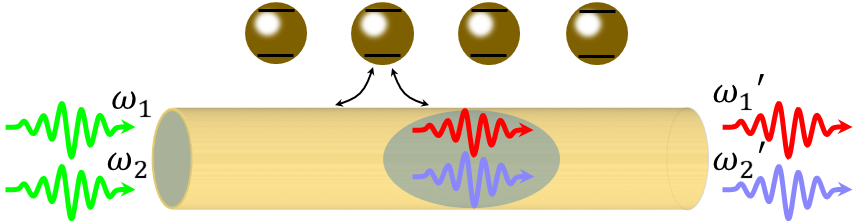}
	\caption{Schematic illustration of the photon pairs propagating along a waveguide with a qubit array and exhibiting inelastic scattering.
	}\label{fig:scheme}
\end{figure}

It is appealing and challenging to exploit the quantum nonlinearities at a few-photon level~\cite{chang2014,Konstantin2016,Roy2017,KimbleRMP2018}. One of the simplest nonlinear quantum processes is the inelastic  scattering of photon pairs. It  exists in  waveguides coupled to a single qubit or qubit arrays, see Fig.~\ref{fig:scheme}, and is sensitive to the  two-photon bound states~\cite{Yudson1984,Shen2007,Fan2007,Baranger2015}. The scattering is greatly enhanced when an incoming or outgoing individual photon excites a single-particle subradiant state~\cite{Baranger2015,Albrecht2019,Baranger2019b}, the concept of multi-excitation subradiant states has been put forward ~\cite{Albrecht2019,Molmer2019,Chang2019}.
It has been predicted that the subradiant mode has a fermionic character and a decay rate with cubic suppression in the number of qubits~\cite{Molmer2019,Chang2019,zhang2019subradiant}.
However, the  role of  collective many-body mechanisms in the enhancement of quantum nonlinear processes remains unclear.
%
%%%%%%%%%%%%%%%%%%%%%%%%%%%%%
%%%%%%%%%%%%%%%%%%%%%%%%%%%%%

In this Letter, we reveal that many-body subradiant states can enhance the incoherent scattering of photon pairs in arrays of two-level qubits supporting long-lived photon-photon correlations.
Specifically, we demonstrate sharp scattering resonances when the energy of the two-particle subradiant state matches the total energy of photon pairs
~\cite{ Yudson1984,Shen2007,Yudson2008,Firstenberg2013,Laakso2014,Fang2014,Xu2015}.
Importantly, considered  resonances are not affected by the  destructive quantum interference known to suppress   two-photon scattering~\cite{Yudson1984,Muthukrishnan2004}. The $N$-particle  subradiant states appear only in periodic arrays with at least $2N$ qubits, e.g., the two-particle state requires at least four qubits, etc. It is also  possible to realize resonant condition for  single- and double-excited subradiant states simultaneously.   We develop a  matrix formulation for the rigorous  Green's function technique valid for an arbitrary arrangement of qubits. This allows us to {\it analytically} identify the role of different double-excited states in the scattering and classify them by the coupling strength. In addition to the double-excited superradiant and subradiant states, we introduce a new concept of {\it twilight state}, which is a  product of single-excited bright and subradiant states. Our results demonstrate that the coupling of light to quantum matter  is far from being fully understood even for the classical Dicke model, and thus this opens a new avenue for manipulating quantum interactions, correlations, and entanglement.

{\it Model}. We consider the system shown schematically in Fig.~\ref{fig:scheme}. It consists of $N$ periodically spaced qubits, coupled to $M$ photons in the one-dimensional waveguide, and it is characterized by the Hamiltonian
\begin{multline}\label{eq:H0}
H=\sum\limits_{k}\hbar\omega_{k}a_{k}^{\dag}a_{k}^{\vphantom{\dag}}
+\sum\limits_{j}\hbar\omega_{0}b_{j}^{\dag}b_{j}^{\vphantom{\dag}}+\frac{\hbar\chi}{2}
\sum\limits_{j}b_{j}^{\dag}b_{j}^{\dag}b_{j}^{\vphantom{\dag}}b_{j}^{\vphantom{\dag}}
\\+\frac{\hbar g}{\sqrt{L}}\sum\limits_{j,k}
(b_{j}^{\dag}a_{k}\e^{\rmi k z_{j}}+b_{j}a_{k}^{\dag}\e^{-\rmi k z_{j}})\:.
\end{multline}
Here, $a_{k}$ are the annihilation operators for the waveguide photons  with the wave vectors $k$ (the corresponding frequencies are given by $\omega_k = c |k|$ with the light velocity $c$), $g$ is the interaction constant, $L$ is the normalization length, and $b_{j}$ are the (bosonic) annihilation operators for the qubit excitations with the frequency $\omega_{0}$, located at the point $z_{j}$. In Eq.~\eqref{eq:H0}, we consider the general case of anharmonic multi-level qubits, the two-level case can be obtained in the limit of large anharmonicity ($\chi \to \infty$) where the multiple occupation is suppressed~\cite{Baranger2013,Poshakinskiy2016}. The photons can be traced out in Eq.~\eqref{eq:H0}, yielding an effective model for describing the excitations in the qubits~\cite{Molmer2019,Suppl},
\begin{align}\label{eq:HM}
\mathcal H=\sum\limits_{i,j}H^{(1)}_{i,j}(\omega_{0})b_{i}^{\dag}b_{j}+\frac{\hbar\chi}{2}
\sum\limits_{j}b_{j}^{\dag}b_{j}^{\dag}b_{j}^{\vphantom{\dag}}b_{j}^{\vphantom{\dag}}\:,
\end{align}
where
\begin{equation}\label{eq:H}
H^{(1)}_{ij}(\omega)\equiv \hbar\omega_{0}\delta_{ij}-\rmi\hbar \Gamma_{0}\e^{\rmi\omega/c|z_{i}-z_{j}|}\:,\quad i,j=1\ldots N\:.
\end{equation}
Hamiltonian~\eqref{eq:H} is non-Hermitian, and it takes into account the radiative losses characterized by the radiative decay rate for a single qubit in a waveguide, $\Gamma_{0}=g^{2}/c$. The interaction between the qubits is  long-ranged  since it is mediated by the photons propagating in the waveguide.
We assume  that the spacing between the qubits is small enough so that the non-Markovian Hamiltonian~\eqref{eq:H} with the phases $(\omega/c)|z_{i}-z_{j}|$ can be replaced by $H^{(1)}_{ij}(\omega_{0})$~\cite{Ivchenko2005}.
From now on, we neglect the non-Markovian effects~\cite{Baranger2013}.

%%%%%%%%%%%%%%%%%%%%%%%%%%
\begin{table}[b!]
	\begin{tabular}{c|c|c|c}
		\hline\hline
		& Superradiant & Twilight & Subradiant \\
		& \includegraphics[width=0.12\textwidth]{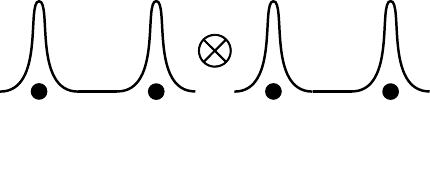} &  \includegraphics[width=0.12\textwidth]{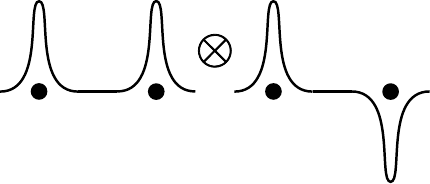} &  \includegraphics[width=0.12\textwidth]{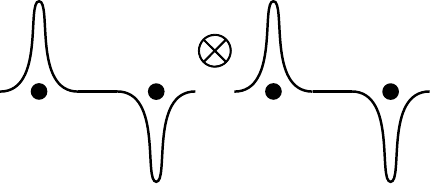} \\
		\hline
		$\sum_j |d_j|^2$ &$\sim N$&$\sim 1 $& $\ll 1$ \\ \hline
		$|\sum_j d_j|^2$ &$\sim N^{2}$&$\ll 1$& $\ll 1$ \\ \hline\hline
	\end{tabular}
	\caption{Classification of the double-excited states depending on the amplitudes of the radiative transition rates $d_j$. Pictures in the upper row sketch the  (nonsymmetrized) two-photon wave function. }\label{tab:twilight}
\end{table}
%%%%%%%%%%%%%%%%%%%%%%%%%%

{\it Double-excited states}. Before proceeding to the study of the scattering of photon pairs, first we analyze double-excited states of the qubit array, $|\Psi\rangle=\sum_{j_{1}j_{2}}\Psi_{j_{1}j_{2}}b_{j_{1}}^{\dag}b_{j_{2}}^{\dag}|0\rangle$.
We can obtain the eigenstates and eigenvalues $2\eps$ by diagonalizing the Hamiltonian Eq.~\eqref{eq:HM}.
We are interested only in the symmetric boson solutions satisfying $\Psi_{i_{1}i_{2}}=\Psi_{i_{2}i_{1}}$.
Due to the qubit-photon interaction, the  double-excited state is unstable, and it will decay into a single-excited state and a freely propagating photon.
The amplidute of the radiative transition from the double-excited state $|\Psi\rangle$ to a single-excited state $b_j^\dag |0\rangle $ is determined by
\begin{eqnarray}
d_{j}&=&\sum\limits_{j'}\e^{\i \omega_0 z_{j'}/c}\Psi_{jj'}. \label{eq:ImEps}
\end{eqnarray}
According to the Fermi's Golden Rule, the total decay rate is given by the sum of the individual decay rates to all single-excited states, and reads
\begin{eqnarray}
\Gamma_1&=&\Gamma_{0} \sum_j |d_j|^2. \label{eq:ImEps}
\end{eqnarray}
Such  decay rate determines  the imaginary part of the eigenvalues, $\Im\eps = -\Gamma_1$. Detailed derivation of Eq.~\eqref{eq:ImEps} is presented in  Supplemental Material~\cite{Suppl}.

The eigenstates are usually classified, depending on a ratio of their  decay rate to that of the individual qubit, as either superradiant ($\Gamma_1 \sim N\Gamma_0$), bright ($\Gamma_1 \sim \Gamma_0$), or subradiant ($\Gamma_1 \ll \Gamma_0$).

However, for double-excited states, this classification is incomplete since it characterizes emission of the first photon only, and it does not provide information about the subsequent emission of the second photon.
Here, we characterize the latter process by the amplitude $\sum_j d_j$, that quantifies the effective dipole moment  of the superposition of single-excited states after emission of the first photon. We identify the states for which the amplitudes of individual radiative transitions are finite but out of phase, so that $\sum_j d_j$ vanishes, as  products of a bright state and a subradiant state, and we term them as {\it twilight states}.
The twilight state quickly decays into a single outgoing photon and a single-excited state. However, the latter excitation appears subradiant and the second photon is emitted after a long time $\sim 1/(\varphi^{2}\Gamma_0)$ with $\varphi\equiv\omega_0|z_2-z_1|/c$, providing long-lived  photon-photon correlations. Namely,  the  correlation function $g^{(2)}(t)$ has contributions with the lifetime $[\sim 1/(\varphi^2\Gamma_0)]$, much longer than that of the individual qubits $(\sim 1/\Gamma_0)$. { The contribution of a twilight state  combines the features  of the subradiant and  bright states. While  it has weak amplitude $\propto \varphi^2$, it can be resonantly excited in a relatively broad spectral range $\sim\Gamma_0$ and decays with a small rate $\sim \varphi^{2}\Gamma_0$.} Detailed analysis  is given in Fig.~S8 in \cite{Suppl}.

\begin{figure}[!t]
	\includegraphics[width=0.5\textwidth]{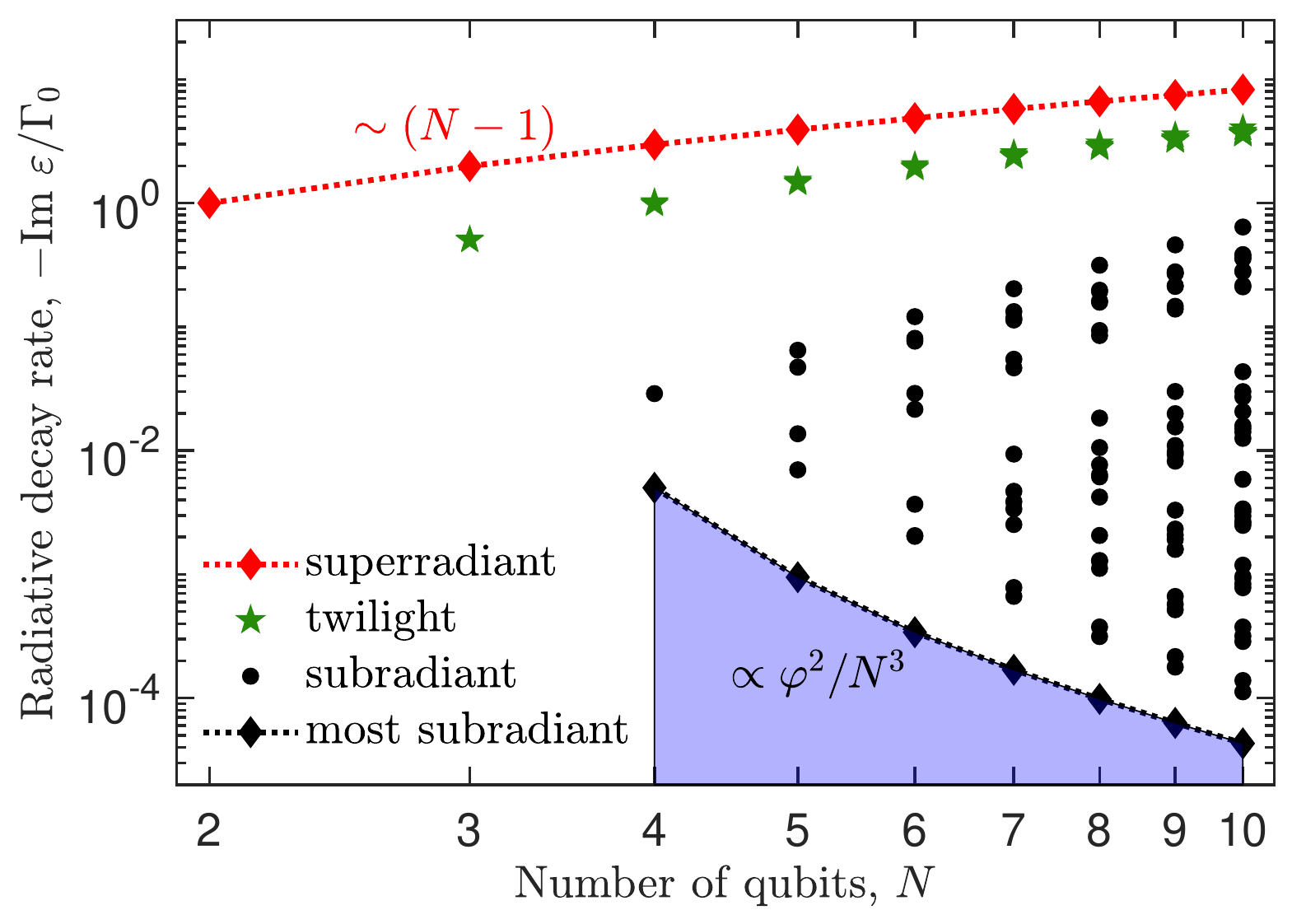}
	\caption{First-order radiative decay rates of double-excited states  depending on the number of qubits in an array $N$.
		Red diamonds, green stars, black dots and black diamonds correspond to the superradiant, twilight, subradiant states and most subradiant states, respectively.
		Calculation has been performed for $\chi=10^4\Gamma_{0},\varphi\equiv \omega_{0}|z_{2}-z_{1}|/c=0.1$.
	}\label{fig:N}
\end{figure}

Thus,  depending on the magnitude of $\sum_j |d_j|^2$ and $|\sum_j d_j|^2$,  the double-excited eigenstates can be classified  as superradiant, twilight and subradiant, see Table~\ref{tab:twilight}.
As demonstrated by our calculations, the short-period array of $N>2$ two-level qubits has one superradiant state, $N(N-3)/2$ subradiant states,  and $(N-1)$ twilight states with total energies around $2\omega_0$.
Figure~\ref{fig:N} shows the dependence of the  decay rate for superradiant states (red diamonds), twilight states (green stars) and subradiant states (black dots and diamonds) on the number of qubits.
For the superradiant state, $\Gamma_1$ is proportional to $(N-1)\Gamma_0$.
For most subradiant  double-excited  state (black diamonds in Fig.~\ref{fig:N}), $\Gamma_1$ becomes smaller by  two orders of magnitude  as the number of qubits increases from $N=3$ to $N=4$, and  for $N\ge 4$ satisfies the scaling relation $\Gamma_1\sim \Gamma_{0}\phi^{2}/N^3$, where $\phi=(\omega_{0}/c)|z_{2}-z_{1}|$.

%%%%%%%%%%%%%%%%%%%%%%%%%%%%%
%%%%%%%%%%%%%%%%%%%%%%%%%%%%%

In order to understand the threshold of $N=4$ qubits for the two-excitation subradiant states, we consider the radiative decay  for the double-excited states in the limiting case where all the qubits are located in the same point, $z_{j}\equiv 0$.
The wavefunction of the subradiant state should satisfy three conditions: (i) $d_{j_1} = \sum_{j_{2}}\Psi_{j_{1}j_{2}}=0$ for all  $j_{1}$, (ii) the symmetricity  $\Psi_{j_{1}j_{2}}=\Psi_{j_{2}j_{1}}$,
and (iii) zero diagonal elements, $\Psi_{j_{1}j_{1}}=0$, since we look for the states where neither of the qubits is occupied twice.  While  these conditions can not be simultaneously met for the
arrays with $N=2$ and $N=3$ qubits,  there exist two subradiant states for $N=4$ qubits with
\begin{equation}\label{eq:sub}
\left[ \Psi_{ij} \right]_1=\tfrac{\sqrt{2}}{4} \left( \begin {smallmatrix} 0&0&1&-1\\0&0&-1&1
\\ 1&-1&0&0\\ -1&1&0&0\end {smallmatrix}
\right)
,\left[ \Psi_{ij} \right]_2=\tfrac{\sqrt{6}}{12} \left( \begin {smallmatrix} 0&-2&1&1\\-2&0&1&1
\\ 1&1&0&-2\\ 1&1&-2&0\end {smallmatrix}
\right),%\nonumber
\end{equation}
where the rows and columns represent the coordinate of the first and second excitation, respectively.
The first state is just  a direct product of the two single-excited subradiant states,
$\tfrac{1}{2}(b_{1}^{\dag}-b_{2}^{\dag})(b_{3}^{\dag}-b_{4}^{\dag})|0\rangle$, while the second state has a more intricate structure.
Due to short length of the array, $N=4$, neither of the subradiant states~\eqref{eq:sub} is described by the fermionic ansatz~\cite{Molmer2019}, see  \cite{Suppl} for more details.
When the spacing between the qubits becomes nonzero, $0<\varphi\ll 1$, these subradiant states become slightly bright: %
\begin{align}\label{eq:eps1}
\eps_1&=\omega_{0}-\phi \Gamma_{0}-\frac{\rmi \phi^{2}}{2} \Gamma_{0},\\\nonumber
\eps_2&=\omega_{0}-\frac{7\phi}{3} \Gamma_{0}-\frac{157\rmi \phi^{2}}{54} \Gamma_{0}\:,
\end{align}
where the first-order decay rates are proportional to $\phi^{2}\ll 1$.
As such, the subradiant states become optically active and can be probed in the light scattering spectra.
More details can be found in~\cite{Suppl}.

{\it Incoherent scattering of photon pairs.} Next, we discuss how the photon-photon interactions are affected by the double-excited states.
To this end we consider the incoherent scattering process, where the two incident photons with the energies  $\omega_1$ and $\omega_2$ are scattered inelastically and converted into  a pair of photons with the energies $\omega_{1}'$ and $\omega_{2}'$, so that $\omega_{1}+\omega_{2}=\omega_{1}'+\omega_{2}'=2\eps$.
Generally, calculation of the scattering is significantly more challenging than that of the double-excited excitations. The reason is that, instead of the reduced problem Eq.~\eqref{eq:HM} describing only the qubit excitations, one needs to consider the full two-particle Hilbert space. Here, we use the rigorous Green function approach, based on the Hamiltonian Eq.~\eqref{eq:H0} with general qubit anharmonicity $\chi$. While our  methodology  is conceptually similar to that of Ref.~\cite{Baranger2013}, it has the advantage of a compact  matrix  formulation valid for arbitrary spatial arrangement of the qubits. Thus, contrary to other Green-function-based techniques~\cite{Kocabas2016,Schneider2016}, we are able to obtain a closed-form  analytical answer. Namely, the $S$-matrix describing the forward incoherent scattering reads
\begin{align}\label{eq:S}
S&(\omega_1',\omega_2';\omega_1,\omega_2) =2\pi \rmi M\delta(\omega_1+\omega_2-\omega_1'-\omega_2'),\\ &M=
-2\rmi\Gamma_0^2{ \,\left(\frac{c}{L}\right)^2}\sum_{i,j} s^{-}_i(\omega_1')s_i^{-}(\omega_2') Q_{ij} s^{+}_j(\omega_1)s^{+}_j(\omega_2)\nonumber
\end{align}
where $s_i^{\pm} = \sum_j G_{ij} \e^{\pm\rmi \omega z_j/c}$ is the structure factor for individual incoming (outgoing) photons,  $ G(\omega)=[\omega-H^{(1)}(\omega)]^{-1}$ is the single-particle Green function, $L$ is the normalisation length. Here, $Q_{ij}$ is the scattering kernel given by $ Q=-\rmi \chi (1-\rmi\chi \Sigma)^{-1}$, where
$
\Sigma_{ij}(\varepsilon) = \int  G_{ij}(\omega)G_{ij}(2\varepsilon-\omega) \rmd\omega/(2\pi) \,.
$
Eq.~\eqref{eq:S} remains valid in the limit of two-level qubits, $\chi\to \infty$, when the $Q\to \Sigma^{-1}$.

The result becomes  more transparent when the Green function is evaluated in the Markovian approximation as $ G(\omega)=[\omega-H(\omega_{0})]^{-1}$. The integration over frequency in $
\Sigma_{ij}(\varepsilon)$ can be then carried out  analytically, yielding
\begin{equation}\label{eq:Q1}
Q_{ij}=\rmi \chi\left[\frac{2\eps- H^{(2)}}{H^{(2)}+\mathcal U-2\eps}\right]_{ii,jj}\:,
\end{equation}
see the Supplemental Material~\cite{Suppl} for details.
Here, the effective two-particle Hamiltonian is given by a sum of individual photon Hamiltonians, $
H^{(2)}_{i_{1}i_{2};j_{1}j_{2}}=\delta_{i_{2},j_{2}}H_{i_{1}j_{1}}+\delta_{i_{1},j_{1}}H_{i_{2}j_{2}}$,
and interaction term
$
\mathcal U_{i_{1}i_{2};j_{1}j_{2}}=\delta_{i_{1}i_{2}}\delta_{j_{1}j_{2}}\delta_{i_{1}j_{1}}\chi\:.
$ The difference $2\eps- H^{(2)}$ in the numerator of Eq.~\eqref{eq:Q1} reflects the destructive quantum interference in the two-photon scattering~\cite{Yudson2008,Muthukrishnan2004}. The  matrix $Q$ has resonances at the eigenstates  $2\eps$  of  the Hamiltonian  Eq.~\eqref{eq:HM} in the two-excitation subspace $H^{(2)}+\mathcal U$. In the  vicinity of the resonance, $\eps\approx \Re\eps_{\nu}$, Eq.~\eqref{eq:Q1} can be simplified to
\begin{equation}\label{eq:Qres}
Q_{ij}(\eps)\approx\frac{2\rmi\Gamma_{0}^{2} d_{i}d^{*}_{j}}{\Re\eps_{\nu}-\rmi\Gamma_{0}\sum_{j'}|d_{j'}|^{2}-\eps}\:,
\end{equation}
where we assume $\chi\to\infty$. The analytical structure of the two-photon kernel $Q$ is now quite clear. The amplitudes of the radiative transitions $d_j$ determine both the resonance linewidth in the denominator [which matches the  decay rate Eq.~\eqref{eq:ImEps}] and the effective oscillator strength of the two-photon resonance in the numerator of Eq.~\eqref{eq:Qres}. This results in the condition $-2\Gamma_{0}\Re\Tr Q=|\Tr Q|^{2}$ that  generalizes  the optical theorem to the interacting two-photon case.

%%%%%%%%%%%%%%%%%%%%%%%%%%%%%
\begin{figure}[t]
	\includegraphics[width=0.46\textwidth]{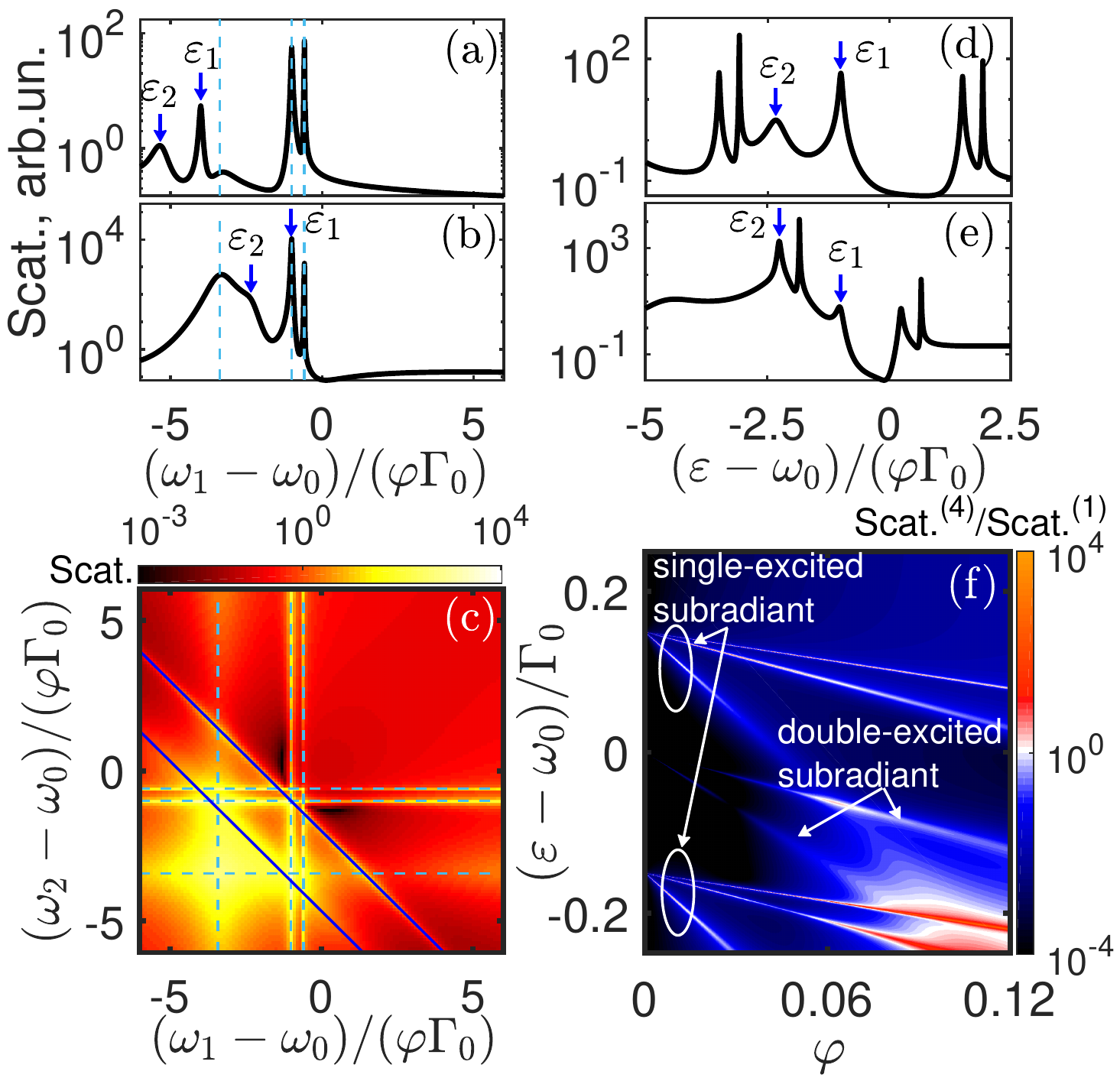}
	\caption{Incoherent forward scattering intensity for an array of four qubits. Scattering intensity as a function of $\omega_1$ for (a) $\omega_{2}-\omega_{1}=6\varphi\Gamma_0$ and (b) $\omega_{2}-\omega_{1}=0$. Thin-dashed vertical lines indicate the positions of single-excited eigenmodes. The arrows show  double-excited subradiant modes  with the energies $\varepsilon_{1,2}$. (c) False color map of the  scattering vs. $\omega_{1}$ and $\omega_2$. Dashed and solid lines indicate the one- and two-photon resonances, respectively. The other parameters in the left panel are $\chi=10^{4}\Gamma_{0},\varphi\equiv \omega_{0}|z_{2}-z_{1}|/c=0.1$.
		(d) and (e): Scattering intensity as a function of average energy $\varepsilon$ for given $\varphi=0.06$ and $\varphi=0.12$, respectively. (f) Normalized false color map of the  scattering vs. array period $\varphi$ and mean energy of incoming photons $\varepsilon$. The other parameters  are  $\chi=10^4\Gamma_0$ and $\omega_2-\omega_1=0.3\Gamma_0$.}\label{fig:scat}
\end{figure}
%%%%%%%%%%%%%%%%%%%%%%%%%%%%%

The calculated incoherent scattering spectra  are summerized in Fig.~\ref{fig:scat}. We present the total forward scattering rate,
\begin{equation}
I(\omega_{1},\omega_{2})=\frac12 \int|M(\omega_1',\omega_1+\omega_2-\omega_1';\omega_1,\omega_2)|^{2} \frac{\rmd \omega_{1}'}{2\pi } \:,
\end{equation}
integrated over the frequencies of the scattered photons.
As such, the scattering map of Fig.~\ref{fig:scat}(c) shows both the resonances when either $\omega_{1}$ or $\omega_{2}$ are tuned to the single-excited subradiant eigenstates
(horizontal and vertical dashed lines), and the two-photon resonances, when the total energy $\omega_{1}+\omega_{2}$ is in resonance with the double-excited subradiant state (diagonal solid lines).
We show the scattering as a function of $\omega_1$ by fixing $\omega_2-\omega_1=6\varphi \Gamma_0$ and $\omega_2-\omega_1=0$, see Fig.~\ref{fig:scat}(a) and \ref{fig:scat}(b), respectively.
Two resonant peaks marked by blue arrows are the positions of $\varepsilon_1$ and $\varepsilon_2$, the energies of double-excited subradiant states.
The outgoing photon pairs can also have strong spatial correlations depending on the nature of the resonant states~\cite{Suppl}.
In the considered case of 4 qubits, there is a point in Fig.~\ref{fig:scat}(c) where the vertical, horizontal and diagonal lines cross over. This  triple-resonant condition occurs when one of the single-excited subradiant states has energy with the same real part  $\omega_{0}-\phi \Gamma_{0}$ as that for the double-excited subradiant state~\eqref{eq:eps1}.
Thus, when both incident photons have the same energies, $\omega_{1}=\omega_{2}=\omega_{0}-\phi \Gamma_{0}$, the triple-resonance  further enhances the scattering, see the peak at $\eps_1$ in Fig.~\ref{fig:scat}(b).

Figures \ref{fig:scat}(d--f) show the  scattering depending  on the  array period $\phi\equiv\omega_0|z_2-z_1|/c$. The spectra are calculated for fixed detuning  $\omega_2-\omega_1=0.3\Gamma_0$ and are normalized to the maximum of total forward scattering for a single qubit, $N=1$.  Dark region around   $\eps=\omega_0$, $\varphi=0$ in  Fig.~\ref{fig:scat}(f) reflects that the scattering is suppressed by the destructive  quantum interference~\cite{Yudson1984,Muthukrishnan2004}, which also exists in $N=2,\ 3$ qubits~\cite{Suppl}.  However, due to subradiant resonances emerging for $\varphi>0$,  the scattering for $N=4$  can exceed that for $N=1$  by several orders of magnitude. The single- and double-excited subradiant resonances can be traced by their different dependence on $\phi$. 
The scattering reaches a local maximum at double resonant conditions.
Namely,  either a single-excited and a double-excited resonance or two single-excited resonances  can occur simultaneously, see the spectra in Fig.~\ref{fig:scat}(d,e), respectively.

%%%%%%%%%%%%%%%%%%%%%%%%%%
{\it Multi-excited states.}
%%%%%%%%%%%%%%%%%%%%%%%%%%
The considered subradiant states are not limited to double excitations. We expect even richer physics for the  excitations with a higher number of photons, $M>2$, which can already be accessed experimentally~\cite{Liang2018}.
As $M$ increases, a threshold of the qubit number for subradiant states will also changes.
To reveal how the subradiant state depends on the excitation number $M$ and qubit number $N$, we find the eigenstate with the energy around $M\omega_0$ that has the minimal  decay rate for different $M$ and $N$~\cite{Suppl}.
The threshold, determined by decrease of $\Gamma_1$ down to $\sim \varphi^2$, occurs for $N=2M$.
The notion of metastable twilight states can also be  extended in the general $M$-body case:  $M'$ particles in bright (or even superradiant) states multiplied by $M-M'$ subradiant states.
How these multi-excited states affect the incoherent $M$-photon scattering is beyond the scope of this Letter.

{\it Conclusion.} We believe that our results open a new research direction for harnessing light-matter interactions in quantum photonics. In particular, the subradiant states boost the incoherent scattering while the
twilight  states  perpetuate the photon-photon correlations.  Custom-tailored long-lived entangled photons could be employed for storage and processing of quantum information.

%%%%%%%%%%%%%%%%%%%%%%%%%%%%%%%%%%%%%%%%%%%%%%%%%

\begin{acknowledgments}
	We acknowledge useful discussions with J. Brehm, I.~Iorsh, A.V. Kavokin, E. Redchenko,  A.A. Sukhorukov, A.V.~Ustinov, and V.I.~Yudson. This work was supported by the Australian Research Council. C. Lee was supported by the National Natural Science Foundation of China (NNSFC) (grants
	11874434 and 11574405). Y.~Ke was partially supported by the International Postdoctoral Exchange Fellowship Program (grant 20180052). A.V.P. also acknowledges a partial support from the Russian President Grant No. MK-599.2019.2 and the Foundation ``BASIS''.
\end{acknowledgments}


%%%%%%%% Supplementary Material %%%%%%%%
\onecolumngrid
\clearpage

\renewcommand {\Im}{\mathop\mathrm{Im}\nolimits}
\renewcommand {\Re}{\mathop\mathrm{Re}\nolimits}
\renewcommand {\i}{{\rm i}}
\renewcommand {\phi}{{\varphi}}
\renewcommand {\rmi}{{\rm i}}
\renewcommand {\rmd}{{\rm d}}
\renewcommand {\sign}{\mathop{\mathrm{sign}}\nolimits}
\renewcommand {\e}{{\rm e}}
\renewcommand {\eps}{\varepsilon}
\renewcommand{\done}{\marginpar{\checkmark}}
\renewcommand {\rot}{\mathop\mathrm{rot}\nolimits}
\renewcommand {\Tr}{{\rm Tr}\,}

\begin{center}
	\noindent\textbf{\large{Supplemental Material:}}
	\\\bigskip
	\noindent\textbf{\large{Inelastic scattering of photon pairs in qubit arrays with subradiant states}}
	\\\bigskip
	\onecolumngrid
	
	Yongguan Ke$^{1,2}$, Alexander V. Poshakinskiy$^{3}$, Chaohong Lee$^{1,4,*}$,  Yuri S. Kivshar$^{2,5}$, Alexander N. Poddubny$^{2,3,5,\dag}$
	
	%\\\vspace{0.1cm}
	\small{$^1$ \emph{Laboratory of Quantum Engineering and Quantum Metrology, School of Physics and Astronomy,
			Sun Yat-Sen University (Zhuhai Campus), Zhuhai 519082, China}}\\
	\small{$^2$ \emph{Nonlinear Physics Centre, Research School of Physics, Australian National University, Canberra ACT 2601, Australia}}\\
	\small{$^3$ \emph{Ioffe Institute, St. Petersburg 194021, Russia}}\\
	\small{$^4$ \emph{State Key Laboratory of Optoelectronic Materials and Technologies, Sun Yat-Sen University (Guangzhou Campus), Guangzhou 510275, China}}\\
	\small{$^5$ \emph{ITMO University, St. Petersburg 197101, Russia}}
\end{center}

\setcounter{equation}{0}
\newcounter{sfigure}
\setcounter{sfigure}{1}
\setcounter{table}{0}
\renewcommand{\theequation}{S\arabic{equation}}

\renewcommand\thefigure{S{\arabic{figure}}}
\renewcommand{\thesection}{S\arabic{section}}
%\renewcommand{\thesubsection}{S\arabic{subsection}}
% \renewcommand{\bibnumfmt}[1]{[S#1]}
%\tableofcontents

\section{S1. Effective model for the excitations}
In this section, we derive the effective Hamiltonian describing the motion of excitations in the qubits.
The tunneling of excitation between different qubits are mediated by the emission and absorption of a photon.
%
%For the excitation with energy $\omega$ initially at $j$th qubit hopping to the $i$th qubit, we assume the energy difference between initial and final state is $\omega$.
%
The hopping amplitude of excitation with energy $\omega$ from $j$-th to $i$-th qubit is given by
\begin{eqnarray}
H_{i,j}^{(1)}&=&\omega_0\delta_{i,j}+g^2 \sum_{l,l'}\int\frac{d k }{2\pi}\e^{\i (z_l-z_l')}\frac{\langle 0|b_i a_k b_l^\dag b_l' a_k^{\dag}  b_j^\dag|0\rangle }{\omega-\omega_k+{0^+}\i} \nonumber \\
&=&\omega_0\delta_{i,j}+g^2 \sum_{l,l'}\int\frac{d k }{2\pi}\frac{\e^{\i (z_i-z_j)}}{\omega-c|k|+{0^+}\i} \nonumber \\
&=& \omega_0\delta_{i,j}-\i\frac{g^2}{c}\e^{\i \omega/c |z_i-z_j|}
\end{eqnarray}
Here, we have used the Cauchy integral formula. We define $\Gamma_0={g^2}/{c}$ as the radiative decay rate.
Then, the total effective Hamiltonian is given as
\begin{align}
\mathcal H=\sum\limits_{i,j}H^{(1)}_{i,j}(\omega_{0})b_{i}^{\dag}b_{j}+\frac{\chi}{2}
\sum\limits_{j}b_{j}^{\dag}b_{j}^{\dag}b_{j}^{\vphantom{\dag}}b_{j}^{\vphantom{\dag}}\:.
\end{align}
When being limited to the subspace with only two excitations,  we can construct the effective two-photon Hamiltonian
\begin{equation}\label{eq:H2U}
H^{(2)}+\mathcal U
\end{equation}
where
\begin{equation}
H^{(2)}=H^{(1)}\otimes I+I\otimes H^{(1)},\quad
\end{equation}
is the sum of individual Hamiltonians for first and second photon, where $\otimes$ denotes the direct product. Explicitly,
\begin{equation}
H^{(2)}_{i_{1}i_{2};j_{1}j_{2}}=\delta_{i_{2},j_{2}}H^{(1)}_{i_{1}j_{1}}+\delta_{i_{1},j_{1}}H^{(1)}_{i_{2}j_{2}},\quad
i_{1},i_{2},j_{1},j_{2}=1\ldots N.
\end{equation}
The Hamiltonian $\mathcal U$ describes the interaction term part,
\begin{equation}
\mathcal U_{i_{1}i_{2};j_{1}j_{2}}=\delta_{i_{1}i_{2}}\delta_{j_{1}j_{2}}\delta_{i_{1}j_{1}}\chi\:.
\end{equation}
The linear eigenvalue problem to obtain the two-particle excitations then reads
\begin{equation}\label{eq:2}
(H^{(2)}+\mathcal U)\Psi=2\eps \Psi\:.
\end{equation}
We do need not all $N^{2}$ solutions of Eq.~\eqref{eq:2} but only the solutions symmetric with respect to the permutation of 1-st and 2-nd photons, i.e. only the bosonic states.

\section{S2. Decay process}
The cascade decay process can be simply decomposed as two processes: (i) two-excitation eigenstate $|\nu\rangle$ decays into one photon and one excitation state with decay rate $\Gamma_{21}^{\nu}$, and (ii) the one excitation eigenstate $|\mu\rangle$ decays into another photon with decay rate $\Gamma_{10}^{\mu}$.
Because there are $N$ one-excitation state $|\mu\rangle$, the first decay process has $N$ decay channels.
We assume that the probability for the decay channel $|\nu\rangle\rightarrow |\mu\rangle$ is $D_{\nu,\mu}$.
Before understanding the whole cascade process, we first show how to calculate the decay rate $\Gamma_{21}^{\nu}$, $\Gamma_{10}^{\mu}$ and probability $D_{\nu,\mu}$.
%Since the second decay

\subsection{A. Decay rate}
We assume the double-excited eigenstate  as $\left|\nu\right\rangle=\sum_{j,m}\psi_{j,m}\left|j,m\right\rangle$. The decay rate of eigenstate $\left|\nu\right\rangle$ can be directly calculated by using the Fermi Golden rule:
\begin{align}\label{eq:SG1}
\Gamma_{21}^{\nu}&=\pi\sum\limits_{k,j}\delta(\omega_{k}+\omega_{0}-2\Re\eps_{\nu})
|\langle 0|a_{k} b_{j}H|\nu\rangle|^{2}\\ \nonumber
&=
g^2\sum\limits_{k,j}\delta(\omega_{k}+\omega_{0}-2\Re\eps_{\nu})
\Bigl|\sum\limits_{m}\e^{\rmi k z_{m}}\Psi_{jm}\Bigr|^{2} = \Gamma_{0}\sum\limits_{j}|d_{j}|^{2},
\end{align}
where
\begin{eqnarray}
d_{j}&=&\sum\limits_{j'}\e^{\i \omega_0 z_{j'}/c}\Psi_{jj'}.
\end{eqnarray}
Using the identity
\begin{equation}
\pi\delta(\omega_{k}+\omega_{0}-2\Re\eps_{\nu})=\Im \frac1{\omega_{k}+\omega_{0}-2\Re\eps_{\nu}-\rmi 0},
\end{equation}
we can also rewrite radiative decay rate as
\begin{equation}\label{eq:imeps}
\Gamma_{21}^{\nu}=\Gamma_{0}\Re\sum\limits_{j,m,m'}\Psi_{jm}\Psi^{*}_{jm'}\e^{\rmi \omega_{0}|z_{m}-z_{m'}|/c}\:.
\end{equation}
The same result could be obtained by just using the fact that $\Gamma_{21}^{\nu}$ is equal to $-\Im \eps_{\nu}$ and $\eps_{\nu}$ is the eigenvalue of the problem Eq.~\eqref{eq:2}.

Similarly, the  decay rate of the single-excited eigenstate $\left|\mu\right\rangle=\sum_{j}\psi_{j}\left|j\right\rangle$.
is given by
\begin{eqnarray}
\Gamma_{10}^{\mu}=\Gamma_{0}\Re\sum\limits_{j,j'}\psi_{j}\psi^{*}_{j'}\e^{\rmi \omega_{0}|z_{j}-z_{j'}|/c}=-\Im\eps_{\mu},
\end{eqnarray}
where $\eps_{\mu}$ is the eigenvalue of the Hamiltonian $H^{(1)}$.

At last, we show the probability for the decay from $\left|\nu\right\rangle$ to $\left|\mu\right\rangle$.
After emission of one photon, the double-excited state $\left|\nu\right\rangle$ is transferred to $\left|\nu'\right\rangle =\sum_{j,m}\Psi_{jm} \e^{\rmi\omega_{0}z_{m}/c} \left|j\right\rangle$.
The probability $D_{\nu,\mu}$ is just related to the overlap between state $\left|\nu'\right\rangle$ and $\left|\mu\right\rangle$, that is,
\begin{eqnarray}
D_{\nu,\mu}=\frac{|\langle \nu' |\mu\rangle|^2}{\sum_j |d_j|^2}=\frac{\Gamma_0}{\Gamma_{21}^{\nu}}{|\langle \nu' |\mu\rangle|^2},
\end{eqnarray}
where $\sum_j |d_j|^2$ is a normalization factor.

\subsection{B. Cascade decay as a function of time}

\begin{figure}[!h]
	\centering\includegraphics[scale=0.8]{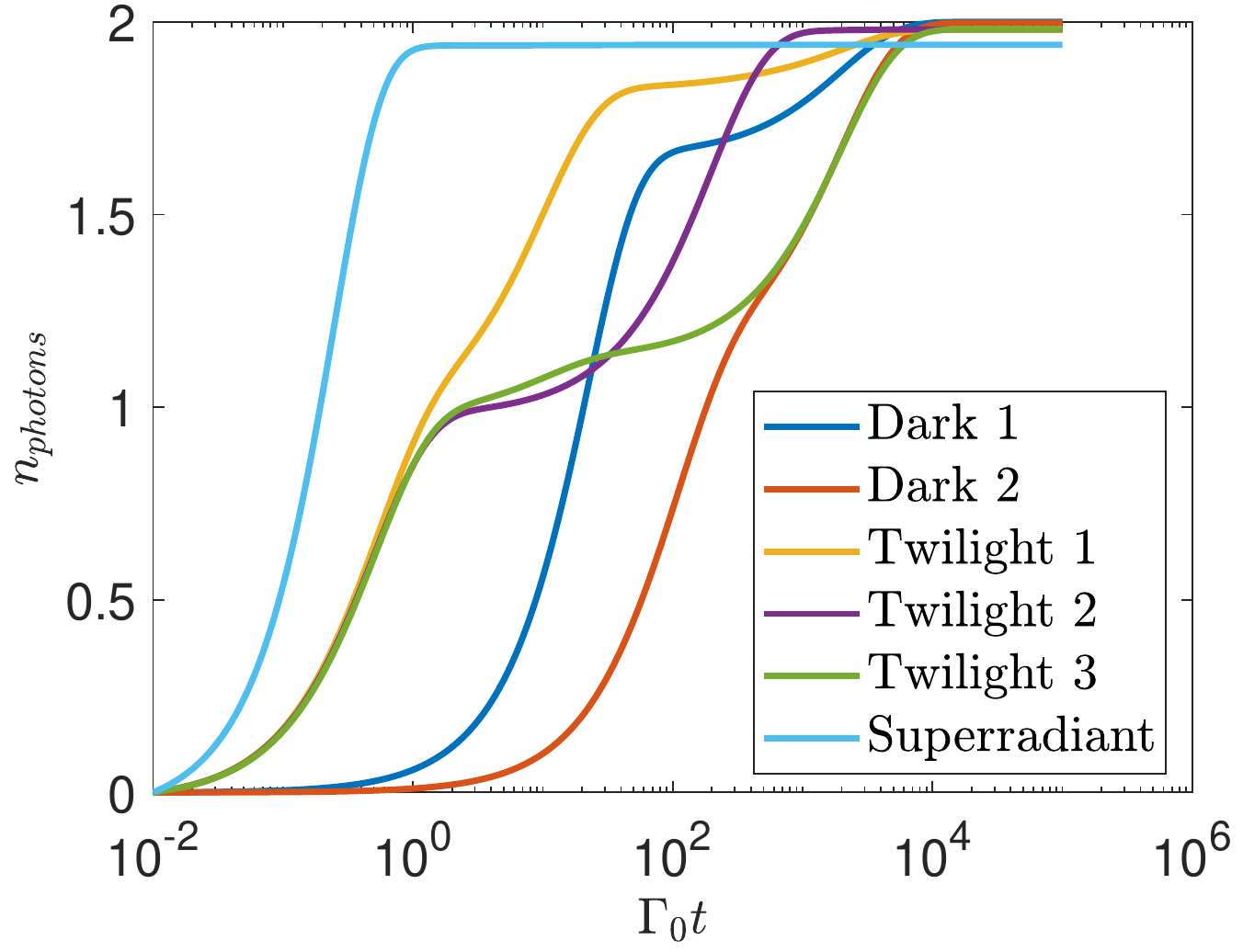}
	\caption{The number of emitted photons as a function of time for different initial states. The initial states are chosen as the double-excited eigenstates, and the parameters are chosen as $N=4$, $\phi=0.1$ and $\chi=10^4\Gamma_0$.	 }\label{DecayTime}
\end{figure}
%The emission of photons is governed by the kinetic equation.
We define the probability of existence of two-excitation state $|\nu\rangle$ and one-excitation state $|\mu\rangle$ at time $t$ as $P_2^{\nu}(t)$ and $P_1^{\mu}(t)$, respectively.
%
%The probability of the two-excitation state $|\mu\rangle$ decay into one-excitation state $|\nu\rangle$  is $D_{\mu,\nu}$.
%
The relation between $P_2^{\nu}(t)$ and $P_1^{\mu}(t)$ satisfies the general kinetic equation~\cite{Kashcheyevs2016S},
\begin{eqnarray}
\frac{d P_2^{\nu}(t)}{dt} &=& -2\Gamma_{21}^{\nu} P_2^{\nu}(t), \label{cascade1} \\
\frac{d P_1^{\mu}(t)}{dt} &=& -2\Gamma_{10}^{\mu} P_1^{\mu}(t)+2\Gamma_{21}^{\nu} D_{\nu,\mu} P_2^{\nu}(t) \label{cascade2}.
\end{eqnarray}
with initial condition $P_2^{\nu}(0)=1$ and $P_1^{\mu}(0)=0$. Because the decay rate characterizes the decay of wave function, an additional factor $2$ should be here for the decay of probability.
Solving Eq.~\eqref{cascade1}, one can obtain
\begin{eqnarray}
P_2^{\nu}(t)=\e^{-2 \Gamma_{21}^{\nu}t}. \label{P2t}
\end{eqnarray}
Substituting Eq.~\eqref{P2t} into Eq.~\eqref{cascade2}, one can obtain
\begin{eqnarray}
\frac{d P_1^{\mu}(t)}{dt} &=& -2\Gamma_{10}^{\mu} P_1^{\mu}(t)+2\Gamma_{21}^{\nu} D_{\nu,\mu} \e^{-2\Gamma_{21}^{\nu}t}.
\end{eqnarray}
We assume $P_1^{\mu}(t)=C^{\mu}(t)\e^{-2\Gamma_{10}^{\mu}t}$, where $C^{\mu}(t)$ satisfies
\begin{eqnarray}
\frac{d C^{\mu}(t)}{dt}=2\Gamma_{21}^{\nu}D_{\nu,\mu}\e^{-2(\Gamma_{21}^{\nu}-\Gamma_{10}^{\mu})t}.
\end{eqnarray}
with boundary condition $C^{\mu}(0)=0$.
It is clear that $C^{\mu}(t)$ is given as
\begin{eqnarray}
C^{\mu}(t)=\frac{\Gamma_{21}^{\nu}D_{\nu,\mu}}{\Gamma_{10}^{\mu}-\Gamma_{21}^{\nu}}(\e^{-2(\Gamma_{21}^{\nu}-\Gamma_{10}^{\mu})t}-1).
\end{eqnarray}
$P_1^{\nu}(t)$ is finally given as
\begin{eqnarray}
P_1^{\mu}(t)=\frac{\Gamma_{21}^{\nu}D_{\nu,\mu}}{\Gamma_{10}^{\mu}-\Gamma_{21}^{\nu}}(\e^{-2\Gamma_{21}^{\nu }t}-\e^{-2\Gamma_{10}^{\mu}t}).
\end{eqnarray}
The total decay rate into photons at time $t$ is given as
\begin{eqnarray}
\Gamma_{tot}(t)=2\Gamma_{21}^{\nu}P_2^{\nu}(t) +\sum\limits_{\mu} 2\Gamma_{10}^{\mu}P_1^{\mu}(t).
\end{eqnarray}
The total number of emitted photons at time $t$, $n(t)$, can be obtained by numerically solving the following equation
\begin{eqnarray}
\frac{d n(t)}{dt}=\Gamma_{tot}(t).
\end{eqnarray}
Fig.~\ref{DecayTime} shows the number of emitted photons for different kinds of initial states.
The parameters are chosen as $N=4$, $\phi=0.1$ and $\chi=10^4\Gamma_0$.	
The decay behaviours of double-excited subradiant, twilight and superradiant states are quite different.
For the superradiant state, both of the two photons are most quickly and simultaneously emitted.
For the subradiant states, both of the two photons are emitted only after much longer time.
For twilight states, the first photon is quickly emitted, and the second photon is emitted after longer time.
The different decay behaviours give a clear classification of the double-excited states.

\section{S3. Subradiant states}
\subsection{A. Eigstates and eigenvalues for $N = 4$ qubits}
In this subsection, we obtain explicit expressions for the subradiant double-excited states  in an array of four two-level qubits with the subwavelength spacing, $\varphi=\omega_{0}d/c\ll 1$.
Since we assume two-level qubits, $\chi\rightarrow \infty $, the double occupation is impossible. Hence, we look for the eigenstates in the following basis,
\begin{eqnarray}
|\psi_{1}\rangle&=&b_{1}^{\dag}b_{2}^{\dag}|0\rangle,\quad |\psi_{2}\rangle=b_{1}^{\dag}b_{3}^{\dag}|0\rangle\:, \nonumber \\
|\psi_{3}\rangle&=&b_{1}^{\dag}b_{4}^{\dag}|0\rangle,\quad |\psi_{4}\rangle =b_{2}^{\dag}b_{3}^{\dag}|0\rangle\:,\nonumber \\
|\psi_{5}\rangle &=&b_{2}^{\dag}b_{4}^{\dag}|0\rangle,\quad |\psi_{6}\rangle=b_{3}^{\dag}b_{4}^{\dag}|0\rangle\:.
\end{eqnarray}
The two-particle Schr\"odinger equation Eq.~\eqref{eq:2} can be expanded in such basis. This is
equivalent to solution of the linear eigenproblem
${\Gamma_0} H |\psi\rangle = {(\varepsilon-\omega_0)} |\psi\rangle$ with the Hamiltonian
\begin{eqnarray}
H= -\frac{1}{2} \left( {\begin{array}{*{20}{c}}
	{{\rm{2i}}}&{{\rm{i}}{e^{{\rm{i}}\varphi }}}&{{\rm{i}}{e^{{\rm{i2}}\varphi }}}&{{\rm{i}}{e^{{\rm{i2}}\varphi }}}&{{\rm{i}}{e^{{\rm{i3}}\varphi }}}&0\\
	{{\rm{i}}{e^{{\rm{i}}\varphi }}}&{{\rm{2i}}}&{{\rm{i}}{e^{{\rm{i}}\varphi }}}&{{\rm{i}}{e^{{\rm{i}}\varphi }}}&0&{{\rm{i}}{e^{{\rm{i3}}\varphi }}}\\
	{{\rm{i}}{e^{{\rm{i2}}\varphi }}}&{{\rm{i}}{e^{{\rm{i}}\varphi }}}&{{\rm{2i}}}&0&{{\rm{i}}{e^{{\rm{i}}\varphi }}}&{{\rm{i}}{e^{{\rm{i2}}\varphi }}}\\
	{{\rm{i}}{e^{{\rm{i2}}\varphi }}}&{{\rm{i}}{e^{{\rm{i}}\varphi }}}&0&{{\rm{2i}}}&{{\rm{i}}{e^{{\rm{i}}\varphi }}}&{{\rm{i}}{e^{{\rm{i2}}\varphi }}}\\
	{{\rm{i}}{e^{{\rm{i3}}\varphi }}}&0&{{\rm{i}}{e^{{\rm{i}}\varphi }}}&{{\rm{i}}{e^{{\rm{i}}\varphi }}}&{{\rm{2i}}}&{{\rm{i}}{e^{{\rm{i}}\varphi }}}\\
	0&{{\rm{i}}{e^{{\rm{i3}}\varphi }}}&{{\rm{i}}{e^{{\rm{i2}}\varphi }}}&{{\rm{i}}{e^{{\rm{i2}}\varphi }}}&{{\rm{i}}{e^{{\rm{i}}\varphi }}}&{{\rm{2i}}}
	\end{array}} \right).
\end{eqnarray}
In the subwavelength case, $\varphi \ll 1$, we can make a Taylor expansion of $\varphi$ around $0$ up to second order and  separate the Hamiltonian as $H=H_0+\varphi H_1+\varphi^2 H_2$, where
\begin{eqnarray}
H_0=  -\frac{\rm{i}}{2}\left( {\begin{array}{*{20}{c}}
	{\rm{2}}&{\rm{1}}&1&{\rm{1}}&{\rm{1}}&0\\
	{\rm{1}}&{\rm{2}}&{\rm{1}}&{\rm{1}}&0&{\rm{1}}\\
	{\rm{1}}&{\rm{1}}&{\rm{2}}&0&{\rm{1}}&{\rm{1}}\\
	{\rm{1}}&{\rm{1}}&0&{\rm{2}}&{\rm{1}}&{\rm{1}}\\
	{\rm{1}}&0&{\rm{1}}&{\rm{1}}&{\rm{2}}&{\rm{1}}\\
	0&{\rm{1}}&{\rm{1}}&{\rm{1}}&{\rm{1}}&{\rm{2}}
	\end{array}} \right),\   H_1=\frac{ 1}{2} \left( {\begin{array}{*{20}{c}}
	{\rm{0}}&1&2&2&{\rm{3}}&0\\
	1&{\rm{0}}&1&1&0&{\rm{3}}\\
	2&1&{\rm{0}}&0&1&2\\
	2&1&0&{\rm{0}}&1&2\\
	{\rm{3}}&0&1&1&{\rm{0}}&1\\
	0&{\rm{3}}&2&2&1&{\rm{0}}
	\end{array}} \right),\ H_2=\frac{{{\rm{i}}}}{4}\left( {\begin{array}{*{20}{c}}
	{\rm{0}}&1&4&4&9&0\\
	1&{\rm{0}}&1&1&0&9\\
	4&1&{\rm{0}}&0&1&4\\
	4&1&0&{\rm{0}}&1&4\\
	9&0&1&1&{\rm{0}}&1\\
	0&9&4&4&1&{\rm{0}}
	\end{array}} \right).
\end{eqnarray}
We treat $H_1$ and $H_2$ as perturbations to $H_0$.
For $H_0$, the eigenstates $(|\Psi_1\rangle, |\Psi_2\rangle,|\Psi_3\rangle,|\Psi_4\rangle,|\Psi_5\rangle, |\Psi_6\rangle)$ and the eigenvalues $(E_1,E_2,E_3,E_4,E_5,E_6)^{T}$ are respectively given as
\begin{eqnarray}
\left( {\begin{array}{*{20}{c}}
	0&{ - \frac{{\sqrt 3 }}{3}}&{ - \frac{1}{2}}&{ - \frac{1}{2}}&0&{\frac{{\sqrt 6 }}{6}}\\
	{\frac{1}{2}}&{\frac{{\sqrt 3 }}{6}}&{ - \frac{1}{2}}&{\frac{1}{2}}&0&{\frac{{\sqrt 6 }}{6}}\\
	{ - \frac{1}{2}}&{\frac{{\sqrt 3 }}{6}}&0&0&{ - \frac{{\sqrt 2 }}{2}}&{\frac{{\sqrt 6 }}{6}}\\
	{ - \frac{1}{2}}&{\frac{{\sqrt 3 }}{6}}&0&0&{\frac{{\sqrt 2 }}{2}}&{\frac{{\sqrt 6 }}{6}}\\
	{\frac{1}{2}}&{\frac{{\sqrt 3 }}{6}}&{\frac{1}{2}}&{ - \frac{1}{2}}&0&{\frac{{\sqrt 6 }}{6}}\\
	0&{ - \frac{{\sqrt 3 }}{3}}&{\frac{1}{2}}&{\frac{1}{2}}&0&{\frac{{\sqrt 6 }}{6}}
	\end{array}} \right), \quad \left( {\begin{array}{*{20}{c}}
	0\\
	0\\
	{{\rm{i}}}\\
	{{\rm{i}}}\\
	{{\rm{i}}}\\
	{3{\rm{i}}}
	\end{array}} \right).
\end{eqnarray}
%and the corresponding ${(2\varepsilon-2\omega_0)}/{\Gamma_0}$ are given as $(0,0,2\i,2\i,2\i,6\i)$.
%
Here, $|\Psi_1\rangle$ and $|\Psi_2\rangle$ are completely dark states, forming a subspace $\mathcal D$. It is instructive to show them as a matrix where indices label coordinates of 1st and 2nd photon.
\begin{equation}
|\Psi_1\rangle=\frac{\sqrt{2}}{4} \left( \begin {smallmatrix} 0&0&1&-1\\0&0&-1&1
\\ 1&-1&0&0\\ -1&1&0&0\end {smallmatrix}
\right)
,|\Psi_2\rangle=\frac{\sqrt{6}}{12} \left( \begin {smallmatrix} 0&-2&1&1\\-2&0&1&1
\\ 1&1&0&-2\\ 1&1&-2&0\end {smallmatrix}
\right) \:. \label{Subradiant12}
\end{equation}
The states
$|\Psi_3\rangle$, $|\Psi_4\rangle$ and $|\Psi_5\rangle$ are the twilight states, see the discussion in the main text.
In particular,  $|\Psi_3\rangle=1/2 (-b_1^\dag+b_4^\dag)\otimes(b_2^\dag+b_3^\dag)|0\rangle$ and $|\Psi_4\rangle=1/2 (b_1^\dag+b_4^\dag)\otimes(-b_2^\dag+b_3^\dag)|0\rangle$ are products of single-excited subradiant and bright state.
$|\Psi_5\rangle$ is entangled state which cannot be decomposed into product form.
%The being an product of single-particle dark and bright state,
%
The state $|\Psi_6\rangle$ is a double-excited superradiant state. The twilight and  superradiant states form a complementary subspace $\mathcal C$.
We respectively define the projector operators upon the subspace $\mathcal D$ and $\mathcal C$ as,
\begin{eqnarray}
P&=&|\Psi_1\rangle\langle \Psi_1|+|\Psi_2\rangle\langle \Psi_2|,\nonumber \\
S&=&\sum\limits_{j\ne 1,2}\frac{1}{E_j-E_1} |\Psi_j\rangle\langle \Psi_j|.
\end{eqnarray}
Applying the degenerate perturbation theory up to $\varphi^2$, the effective Hamiltonian for the subspace $\mathcal D$ is given as
%To seek the dark states under perturbations $H_1$ and $H_2$
\begin{eqnarray}
H_{eff}=\varphi P H_1 P+ \varphi^2( P H_2 P+ P H_1 S H_1 P).
\end{eqnarray}
Since the coupling between twilight states and dark states are negligible, $S$ is simply given as $|\Psi_6\rangle\langle\Psi_6|/(3\i)$.
The eigenvalues of the dark states are approximately given as
\begin{eqnarray}
E_{d, 1}&=&-\phi -\frac{\rmi \phi^{2}}{2} ,\nonumber \\
E_{d,2}&=&-\frac{7\phi}{3} -\frac{157\rmi \phi^{2}}{54}.
\end{eqnarray}
Thus, the corresponding $\eps_{1}$ and $\eps_{2}$ are respectively given as
\begin{align}
\eps_1&=\omega_{0}-\phi \Gamma_{0}-\frac{\rmi \phi^{2}}{2} \Gamma_{0},\\\nonumber
\eps_2&=\omega_{0}-\frac{7\phi}{3} \Gamma_{0}-\frac{157\rmi \phi^{2}}{54} \Gamma_{0}\:,
\end{align}
It is easy to calculate the single-photon dark eigenfrequencies for $N=4$, $\phi\ll 1$, they are,
\begin{equation}
\omega_{d,1}=\omega_{0}-\phi \Gamma_{0}-\frac{\rmi \phi^{2}}{4}\Gamma_{0},\quad
\omega_{d,2}\approx \omega_{0}-0.59 \phi \Gamma_{0}-0.025\rmi \phi^{2}\Gamma_{0},\quad
\omega_{d,3}\approx \omega_{0}-3.4 \phi \Gamma_{0}-5.0\rmi \phi^{2}\Gamma_{0}\:.
\end{equation}
Hence, there can be a double resonance for
\begin{equation}
\eps=\Re\eps_{d,1}=\Re \omega_{d,1}=\omega_{0}-\phi \Gamma_{0}
\end{equation}
%%%%%%%%%%%%%%%%%%%%%%%%%%%%%%%%%%%%%
\subsection{B. Comparison between subradiant states and fermionic ansatz}
\begin{figure}[!htp]
	\centering\includegraphics[scale=0.75]{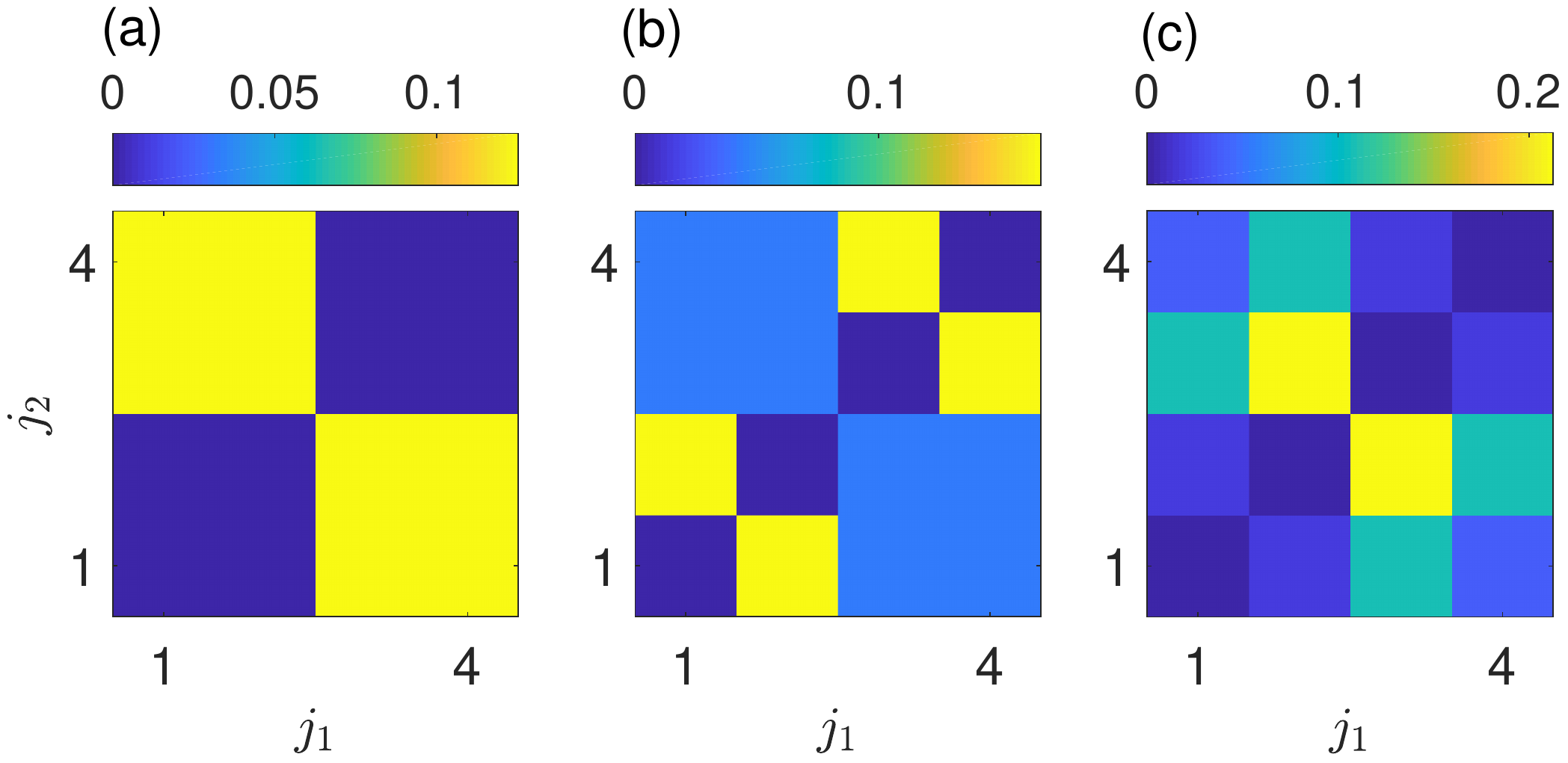}
	\caption{The probability amplitude $|\psi_{j_1,j_2}|^2$ for the most-subradiant state (a), the second most-subradiant state (b) and fermionic state (c) in $N=4$ qubits.}\label{SubradiantN4}
\end{figure}

\begin{figure}[!htp]
	\centering\includegraphics[scale=0.7]{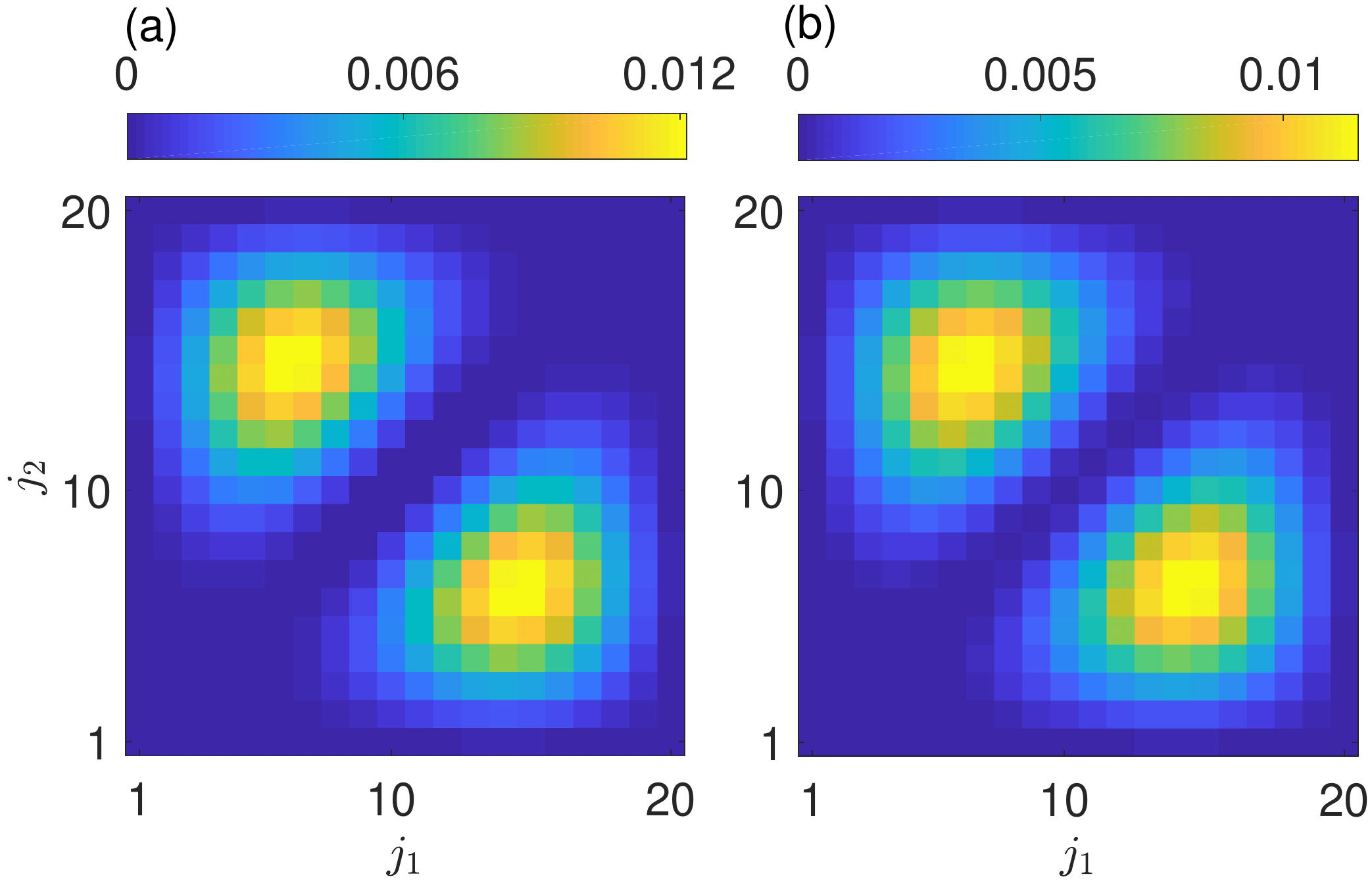}
	\caption{The probability amplitude $|\psi_{j_1,j_2}|^2$ for the most-subradiant state (a) and fermionic state (c) in $N=20$ qubits.}\label{SubradiantN20}
\end{figure}

In Refs.~\onlinecite{Albrecht2019S,Molmer2019S}, the most-subradiant state $|\psi\rangle=\mathcal{N}\sum_{j_1,j_2}\psi_{j_1,j_2}|j_1,j_2\rangle$ with normalized factor $\mathcal N$ is supposed to have form of anti-symmetric combination of single-excitation eigenstates, i.e.,
\begin{eqnarray}
\psi_{j_1,j_2}=\begin{cases}
~c_{j_1}^{\xi=1}c^{\xi=2}_{j_2}-c_{j_1}^{\xi=2}c^{\xi=1}_{j_2},&(j_1 \ge j_2)\\
-(c_{j_1}^{\xi=1}c^{\xi=2}_{j_2}-c_{j_1}^{\xi=2}c^{\xi=1}_{j_2}),&(j_1\le j_2)
\label{ansatz}
\end{cases}
\end{eqnarray}
where $c_{j}^{\xi}$ is amplitude of the single-excited subradiant state $|\xi\rangle=\sum_jc_j^{\xi}|j\rangle$ at the $j$ site.
However, this approximation works well only when the number of qubits $N$ is large enough.
Besides, not all of the subradiant states satisfy the fermionic ansatz~\cite{Molmer2019S,zhang2019subradiantS}.
Here, we compare the probability amplitude of the exact double-excited subradiant state and that of Eq.~\eqref{ansatz} in $N=4, \ 20$ qubits.
The other parameters are chosen as $\phi=0.1$ and $\chi\to \infty$.
In the case of $N=4$ qubits, we calculate probability amplitudes of the two most-subradiant states with energies $\varepsilon_1$ and $\varepsilon_2$ in Fig.~\ref{SubradiantN4} (a) and (b), respectively.
They are very close to the two subradiant states~\eqref{Subradiant12}.
We also calculate the fermionic-like state by anti-symmetric combination of the single-excited subradiant states with energies $\omega_{d,1}$ and $\omega_{d,2}$.
The probability amplitudes are shown in Fig.~\ref{SubradiantN4} (c).
It is clear that the fermionic-like state departs from the exact most-subradiant state in Fig.~\ref{SubradiantN4} (a).
Eq.~\ref{ansatz} is not a good approximation for the most-subradiant state in the small $N$ qubits.
For the second most-subradiant state in Fig.~\ref{SubradiantN4} (b), there is no resemblance by any anti-symmetric combinations of single-excited subradiant states.
In the case of $N=20$ qubits, we show the probability amplitude for both the exact most-subradiant states and fermionic states, see Fig.~\ref{SubradiantN20} (a) and (b), respectively.
It is clear that the Eq.~\eqref{ansatz} can capture the features of the most-subradiant state in large number of qubits.

%%%%%%%%%%%%%%%%%%%%%%%%%%%%%%%%%%%%%%
\section{S4. Two-photon scattering}
\subsection{A. General matrix theory}
\begin{figure}[h]
	\centering\includegraphics[scale=1.5]{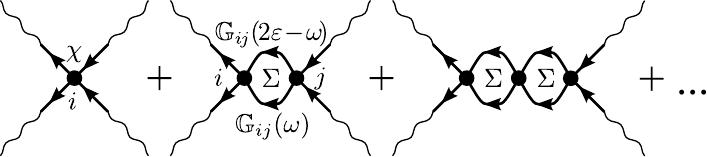}
	\caption{The series corresponding to the calculation of two-photon scattering. Thick lines indicate the qubit Green function Eq.~\eqref{eq:defG}, and the wavy lines are incoming and outgoing photons.}\label{diaU}
\end{figure}

The diagrams corresponding to two-photon scattering are shown in Fig.~\ref{diaU}.
The corresponding amplitude reads~[\onlinecite{Poshakinskiy2016S}]
\begin{eqnarray}\label{S}
&&S(\omega_1',\omega_2';\omega_1,\omega_2)\nonumber \\
&=& \frac{2 g^4}{L^2}\sum_{i,j=1}^{N} s_i^-(\omega_1')s_i^-(\omega_2')\left[ -\rmi \chi \delta_{ij}   + ( -\rmi \chi) \Sigma_{ij} (-\rmi \chi) + \ldots\right] s_j^+(\omega_1)s_j^+(\omega_2)  2\pi\delta(\omega_1+\omega_2-\omega_1'-\omega_2') \nonumber \\
&=&2\pi\rmi M\delta(\omega_1+\omega_2-\omega_1'-\omega_2') ,\quad M= -2\rmi \Gamma_0^2\sum_{i,j=1}^{N} s_i^-(\omega_1')s_i^-(\omega_2')Q_{ij}s_j^+(\omega_1)s_j^+(\omega_2) \,\left(\frac{c}{L}\right)^2
\end{eqnarray}
where  factors $s_i^\pm(\omega)$ describe the external lines of the diagrams,
\begin{align}\label{eq:s}
s_i^\pm(\omega) = \sum_j G_{ij} \e^{\pm\rmi (\omega/c) z_j},
\end{align}
$G_{ij}$ is the Green function for single qubit excitation defined by
\begin{equation}  \label{eq:defG}
(\omega-\omega_0){G}_{ij}(\omega)+\rmi\Gamma_{0}\sum_m \e^{\rmi (\omega/c)|z_i-z_m|} {G}_{mj}(\omega)= \delta_{ij}\:,
\end{equation}
the matrix $Q$ is given by $ Q=-\rmi \chi/(1-\rmi\chi \Sigma)$,
and the matrix $\Sigma$ has the elements
\begin{align}\label{gdef}
\Sigma_{ij}(\varepsilon) = \int  G_{ij}(\omega)G_{ij}(2\varepsilon-\omega) \frac{\rmd\omega}{2\pi} \,,
\end{align}
with $\varepsilon = (\omega_1+\omega_2)/2$.
Eq.~\eqref{S} is equivalent to Eq.~(7) in the main text.
%%%%%%%%%%%%%%%%%%%%%%%%%%%%%%%%%%
\subsection{B. Analytical expansion in the Markovian approximation }
%%%%%%%%%%%%%%%%%%%%%%%%%%%%%%%%%%
We will now restrict ourselves to the Markovian approximation, when the frequency dependence of the phase factors  in Eqs.~\eqref{eq:s} and \eqref{eq:defG} can be neglected and they are
evaluated at the resonant frequency $\omega_{0}$. This is valid in the considered subwavelength regime when $(\omega/c) |z_{i}-z_{j}|\ll 1$. It is then possible to simplify \eqref{S} and to demonstrate, that the resonances in scattering for the total photon energy $\omega_{1}+\omega_{2}=\omega_{1}'+\omega_{2}'$ correspond to the two-particle eigenstates of the Hamiltonian \eqref{eq:H2U}.

We start with noting that
\begin{eqnarray}
\int\frac{\rmd \omega}{2\pi} G_{ij}(\omega)G_{kl}(2\eps-\omega)&=&
\int\frac{\rmd \omega}{2\pi} \left[\frac1{H-\omega}\right]_{ij}\left[\frac1{H+\omega-2\eps}\right]_{kl}
\nonumber \\
&=&\left[\frac{\rmi}{H^{(1)}\otimes I+I\otimes H^{(1)}-2\eps}\right]_{ik,jl}\:.
\end{eqnarray}
Hence,
\begin{equation}\label{eq:Sigma1}
\Sigma_{ij}=\left[\frac{\rmi}{H\otimes 1+1\otimes H-2\eps}\right]_{ii,jj}\equiv \left[\frac{\rmi}{H^{(2)}+\mathcal U-2\eps}\right]_{ii,jj},
\end{equation}
and
\begin{equation}\label{eq:Q1}
Q_{ij}=-\left[\frac{\rmi \chi}{1-\rmi \chi \Sigma}\right]_{ij}=\rmi \chi\left[\frac{2\eps-\mathcal H}{\mathcal H+\mathcal U-2\eps}\right]_{ii,jj}
\end{equation}
Eqs.~\eqref{eq:2}, \eqref{eq:Sigma1} and \eqref{eq:Q1} allow one to calculate two-photon eigenmodes and the scattering spectra using the matrix methods in the Markovian approximation.
We now consider a specific double-excited eigenstate $\Psi$ satisfying
\begin{equation}\label{eq:epsnu}
(\mathcal H+\mathcal U-2\eps_{\nu})\Psi =0
\end{equation}
and expand the $Q$ matrix near $\eps=\eps_{\nu}$.  Our aim is to take the two-level qubit limit $\chi\to\infty$ analytically. We obtain from Eq.~\eqref{eq:Q1}
\begin{equation}
Q_{ij}=\rmi \chi\left[\frac{2\eps-\mathcal H}{\mathcal H+\mathcal U-2\eps}\right]_{ii,jj}\approx
\frac{\rmi \chi^{2} \Psi_{ii} \Psi_{jj}^{*}}{2\eps_{\nu}-2\eps}\:.
\end{equation}
The diagonal matrix elements $\Psi_{ii}\propto 1/\chi$ can be obtained from the wavefunction $\Psi^{(0)}$ calculated in the limit $\chi\to \infty$ by means of the perturbation theory.
Namely,
\begin{equation}
\Psi_{ii}=\frac{2\rmi \Gamma_{0}}{\chi}\sum\limits_{j'}\Psi_{ij'}^{(0)}\e^{\rmi q_{0}|z_{i}-z_{j'}|},
\end{equation}
where we used the condition $\Psi_{ij'}=\Psi_{j'i}$.
Hence, we find
\begin{equation}\label{eq:qan}
Q_{ij}\approx \frac{4\rmi \Gamma_{0}^{2}}{2\eps_{\nu}-2\eps} d_{i}d_{j}^{*},\quad
d_{i}=\sum\limits_{j'}\Psi_{ij'}^{(0)}\e^{\rmi q_{0}|z_{i}-z_{j'}|}\:,
\end{equation}
in agreement with Eq.~(11) in the main text.

We recall that due to radiative decay rate of the two-photon state, its energy has a finite imaginary part
\begin{align}
-\Im \eps_{\nu} = \Gamma_{21}^{\nu}=\Gamma_{0}\sum\limits_{j}|d_{j}|^{2},
\end{align}
see Eq.~\eqref{eq:SG1}.
%It is also useful to consider the  radiative decay rate of the two-photon state. It  can be calculated directly using the Fermi Golden rule:
%\begin{equation}
%\Gamma_{1}=\pi\sum\limits_{k,j}\delta(\omega_{k}+\omega_{0}-2\Re\eps_{\nu})
%|\langle 0|a_{k} b_{j}|H|\nu\rangle|^{2}=
%-\pi\sum\limits_{k,j}\delta(\omega_{k}+\omega_{0}-2\Re\eps_{\nu})
%\Bigl|\sum\limits_{m}\e^{\rmi k z_{m}}\Psi_{jm}\Bigr|^{2}
%\end{equation}
%Using the identity
%\begin{equation}
%\pi\delta(\omega_{k}+\omega_{0}-2\Re\eps_{\nu})=\Im \frac1{\omega_{k}+\omega_{0}-2\Re\eps_{\nu}-\rmi 0}
%\end{equation}
%we obtain the radiative decay rate as
%\begin{equation}\label{eq:imeps}
%\Gamma_{1}=\Gamma_{0}\Re\sum\limits_{j,m,m'}\Psi_{jm}\Psi^{*}_{jm'}\e^{\rmi \omega_{0}|z_{m}-z_{m'}|/c}\equiv \Gamma_{0}\sum\limits_{j}|d_{j}|^{2}\:.
%\end{equation}
%The same result could be also obtained even just using the fact that $\Gamma_{1}$ is equal to $-\Im \eps_{\nu}$ and $\eps_{\nu}$ is the eigenvalue of the problem Eq.~\eqref{eq:epsnu}. Comparing Eq.~\eqref{eq:qan} and Eq.~\eqref{eq:imeps}
Then, we see a connection, the numerator of
$Q$ is proportional to the imaginary part of the denominator. Hence, in the vicinity of the given resonance $\eps_{\nu}$ the following identity holds:
\begin{equation}
-2\Gamma_{0}\Re  \Tr Q=|\Tr Q|^{2}\:.
\end{equation}

\subsection{C. Two photon wave-function}
\begin{figure}[!h]
	\centering\includegraphics[scale=0.7]{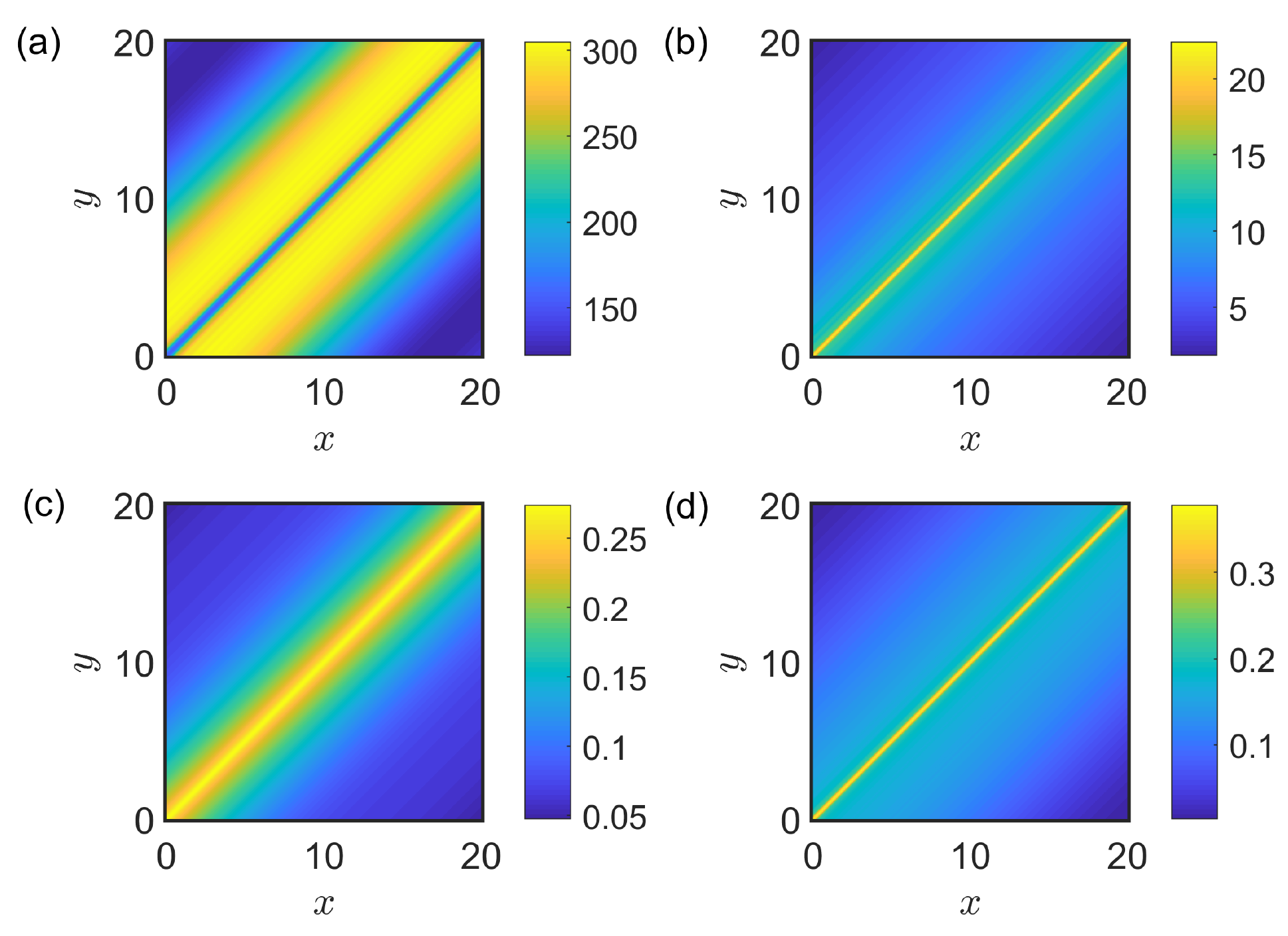}
	\caption{The spatial correlation function of outgoing photon pairs with different input energies: (a) $\omega_{1,2}=\varepsilon_{1}$, (b) $\omega_{1,2}=\varepsilon_{2}$, (c) $\omega_{1,2}=\varepsilon_{1}\mp 3\varphi\Gamma_0$ and (d) $\omega_{1,2}=\varepsilon_{2}\mp 3\varphi\Gamma_0$.
		The other parameters are chosen as $N=4$, $\phi=0.1$ and $\chi=10^4\Gamma_0$. Here, all the sub-figures are in arbitrary unit. }\label{Correlation}
\end{figure}
To reveal the spatial correlation between the two outgoing photons, we make a Fourier transformation of the forward scattering,
\begin{eqnarray}
S(x,y) = \int_{}^{} {S({\omega _{{k_1}'}},{\omega _{{k_2}'}},{\omega _{{k_1}}},{\omega _{{k_2}}})} {e^{i{k_1}'x}}{e^{i{k_2}'y}}d{k_1}'d{k_2}', (x>0,y>0)
\end{eqnarray}
where $x$ and $y$ are the positions of two forward outgoing photons. $|S(x,y)|^2$ indicates the correlation of detecting one photon at $x$ and the other photon at $y$ in the incoherent scattering process.
Fig.~\ref{Correlation} shows the correlation function $|S(x,y)|^2$ of the outgoing photons when the total energies of incoming photon pairs match the first and second double-excited subradiant states.
The parameters are chosen as $N=4$, $\phi=0.1$, $\chi=10^4\Gamma_0$, (a) $\omega_{1,2}=\varepsilon_{1}$, (b) $\omega_{1,2}=\varepsilon_{2}$, (c) $\omega_{1,2}=\varepsilon_{1}\mp 3\varphi\Gamma_0$ and (d) $\omega_{1,2}=\varepsilon_{2}\mp 3\varphi\Gamma_0$, respectively.
These parameters are corresponding to the four resonant peaks of the double-excited subradiant states in Fig.~3(a) and (b) of main text.
When the double-excited subradiant states are excited, the forward outgoing photons show strong spatial correlations.
%\red{Poshakinskiy: \it To claim this, $g^{(2)}$ correlation function should be evaluated. It is different from $S$ by accounting not only incoherent scattering but also coherent transmission. (Anti)bunching is a result of interference of the coherent and incoherent contributions to $g^{(2)}$. }
%
%It means that the outgoing photons are bound states with strong correlations.
%

\subsection{D. Incoherent scattering in two and three qubits}
\begin{figure}[!h]
	\centering\includegraphics[scale=0.8]{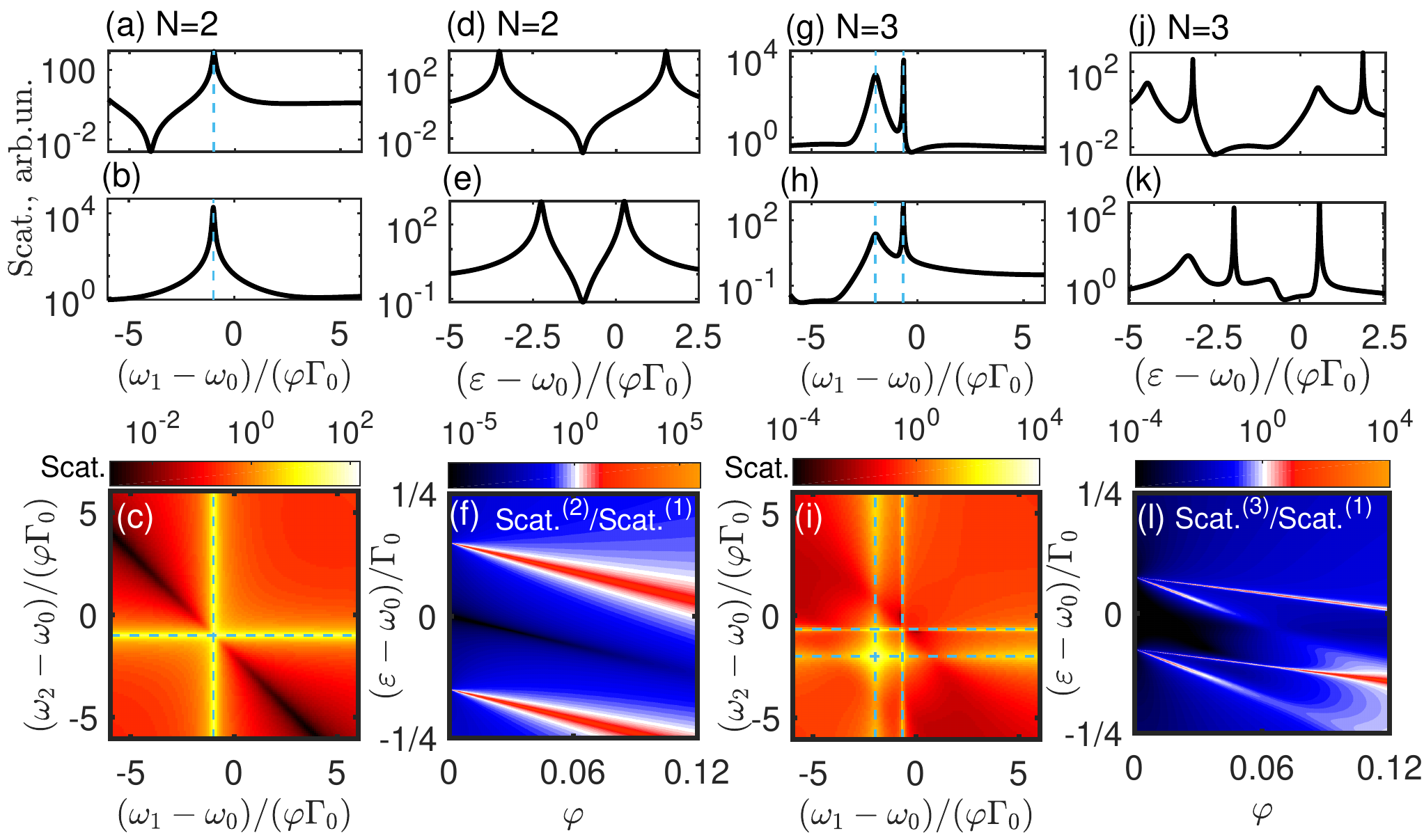}
	\caption{Incoherent forward scattering intensity in two qubits (a)-(f) and three qubits (g)-(l). Scattering intensity as a function of $\omega_1$ for (a): $\omega_{2}-\omega_{1}=6\varphi\Gamma_0$ and $N=2$; (b): $\omega_{2}-\omega_{1}=0$ and $N=2$; (g): $\omega_{2}-\omega_{1}=6\varphi\Gamma_0$ and $N=3$; (h): $\omega_{2}-\omega_{1}=0$ and $N=3$. Thin-dashed vertical lines indicate the positions of single-excited eigenmodes. (c,i): False color map of the  scattering vs. $\omega_{1}$ and $\omega_2$ for (c) $N=2$ and (i) $N=3$. Dashed lines indicate the one-photon resonances. The other parameters in the first and third panels are $\chi=10^{4}\Gamma_{0},\varphi\equiv \omega_{0}|z_{2}-z_{1}|/c=0.1$.
		Scattering intensity as a function of average energy $\varepsilon$ for (d): $\varphi=0.06$ and $N=2$; (e): $\varphi=0.12$ and $N=2$; (j): $\varphi=0.06$ and $N=3$; (k): $\varphi=0.12$ and $N=3$. (f,l): Normalized false color maps of the  scattering vs. $\varphi$ and $\varepsilon$ for (f) $N=2$ and (l) $N=3$. The other parameters in the second and fourth panels are chosen as $\chi=10^4$ and $\omega_2-\omega_1=0.3\Gamma_0$. }\label{ScatN2N3}
\end{figure}
The double-excited subradiant states are present for $N\ge 4$ qubits. Thus, it is instructive to compare the incoherent scattering shown in Fig. 3 in the main text with the scattering in two and three qubits without any double-excited subradiant states, see Fig.~\ref{ScatN2N3}.
In the case of two qubits, there is only one single-excited subradiant state.
The incoherent forward scattering is strongly enhanced when either of the individual incoming photons is resonant with the single-excited subradiant state, see the peaks along the dashed lines in Fig.~\ref{ScatN2N3}(a)-(c),
where the energy differences are fixed as $\omega_2-\omega_1=6\varphi\Gamma_0$ for (a) and $\omega_2-\omega_1=0$ for (b), and other parameters are chosen as $\chi=10^4\Gamma_0$, $\phi=0.1$.
In Fig.~\ref{ScatN2N3}(c), apart from the resonant peaks along the dashed line, there are also dark regions in scattering spectral along the line $\varepsilon-\omega_0=-0.1\Gamma_0$ with $\varepsilon=(\omega_2+\omega_1)/2$.
To better understand the dark regions, we calculate the incoherent scattering vs. $\varepsilon$ and $\varphi$ in Fig.~\ref{ScatN2N3}(f), where the energy difference $\omega_2-\omega_1=0.3\Gamma_0$ and $\chi=10^4$.
The incoherent scattering is normalized by the maximum value of the total incoherent forward scattering in a single qubit.
The incoherent scattering is enhanced when $\rm Scat.^{(2)}/Scat.^{(1)}>1$ and suppressed when $\rm Scat.^{(2)}/Scat.^{(1)}<1$.
It is clear that the dip is lying at $\varepsilon-\omega_0=-\varphi\Gamma_0$.
When $\varphi\equiv \omega_0|z_2-z_1|/c \to 0$, the incoherent scattering is almost negligible due to the destructive quantum interference of two distinct excitation pathways~\cite{Muthukrishnan2004S}.
However, when $\varphi$ increases (i.e. the atoms are not exactly at the same point), fully dark modes become subradiaint. As a result,  more  pathways start playing role in scattering  and not all of these pathways interfere destructively. Thus, the incoherent scattering at the dip increases with $\varphi$.
In Fig.~\ref{ScatN2N3}(d) and (e), we show the incoherent scattering by fixing $\varphi=0.06$ and $\varphi=0.12$, respectively.
It is clear that the incoherent scattering in the dip of Fig.~\ref{ScatN2N3}(e) is larger than that of Fig.~\ref{ScatN2N3}(d).

We do the similar calculations for the case of $N=3$ qubits, where the parameters are chosen the same as the counterparts in the case of two qubits.
Since there are $N-1=2$ single-excited subradiant states for three qubits,
there are two resonant peaks in the incoherent scattering spectral where the energy of the individual incoming photons hits the single-excited subradiant states, see Fig.~\ref{ScatN2N3}(g)-(i).
When the total energy of the incoming photons are closed to $2\varepsilon\approx 2\omega_0$, the incoherent scattering is also strongly suppressed due to the destructive interference, see Fig.~\ref{ScatN2N3}(j)-(l).
Similar to the case of two qubits, there are  incoherent scattering can be enhanced only due to the single-excited subradiant states, and the incoherent scattering in the dip increases with $\varphi$.

As discussed in the main text, Fig.~3, for $N=4$ qubits double-excited subradiant states start playing role. They appear right in the dip with the resonant energy $\propto \varphi$, and further enhance the incoherent scattering.

%%%%%%%%%%%%%%%%%%%%%%%%%%%%%%%%%%%%%%%%%%
\section{S5. Threshold of qubit number for subradiant states}
%%%%%%%%%%%%%%%%%%%%%%%%%%%%%%%%%%%%%%%%%

\begin{figure}[!h]
	\includegraphics[width=0.8\textwidth]{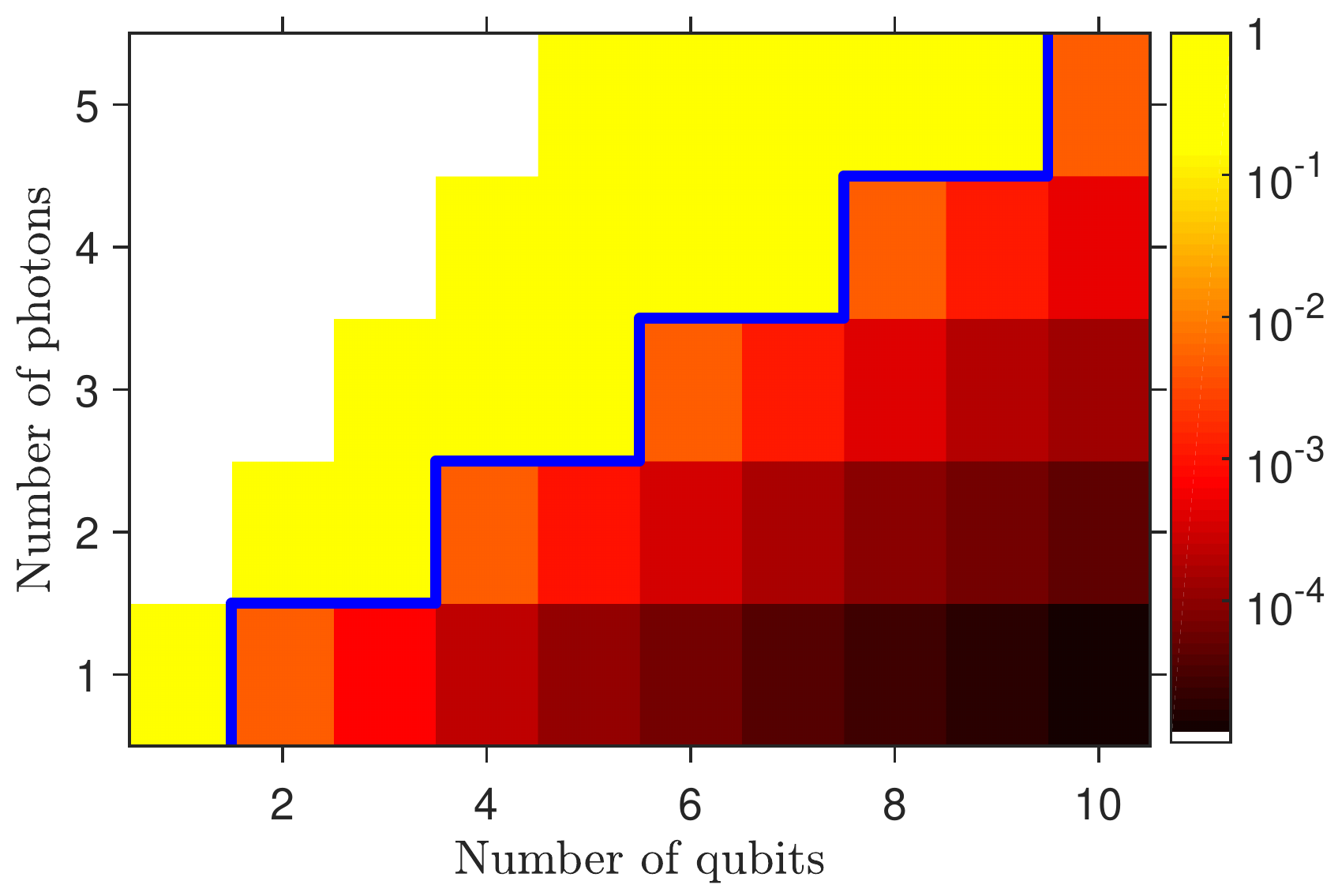}
	\caption{The first-order decay rate as function of the excitation number and qubit number. Calculation has been performed for the parameters $\varphi=0.1$ and $\chi=10^4\Gamma_0$. }\label{threshold}
\end{figure}

In the main text, we show that the double-excited subradiant state appears in the arrays with at least $4$ qubits.
Generally, one may ask how many qubits support the appearance of $M$ excitation subradiant states.
To reveal the threshold of the existence of subradiant state, we find the eigenstate with the energy around $M\omega_0$ that has the minimal first-order decay rate. The decay rate of such state as excitation number $M$ and qubit number $N$ change is shown in Fig.~\ref{threshold}.
The calculation is based on the diagonalization of effective Hamiltonian, Eq.~(2) of the main text, with the parameters $\varphi=0.1$ and $\chi=10^4\Gamma_0$.
The white region is unphysical for two-level qubits, since $M$ excitations cannot occupy the $N<M$ qubits.
The threshold of subradiant state, determined as the moment when the decay rate drops down to $\Gamma_1\sim \varphi^2$, is denoted by the blue solid line.
Thus, we can deduce that the threshold satisfies $N=2M$, in other words,  the $M$-excited subradiant states can be engineered only if the number of qubits exceeds $2M$.

%%%%%%%%%%%%%%%%%%%%%%%%%%%%%%%%%%%%%%%%%%
\section{S6. Subradiant and twilight states in photon-photon correlations}
%%%%%%%%%%%%%%%%%%%%%%%%%%%%%%%%%%%%%%%%%%
%%%%%%%%%%
\begin{figure}[!htp]
	\centering\includegraphics[width=\textwidth]{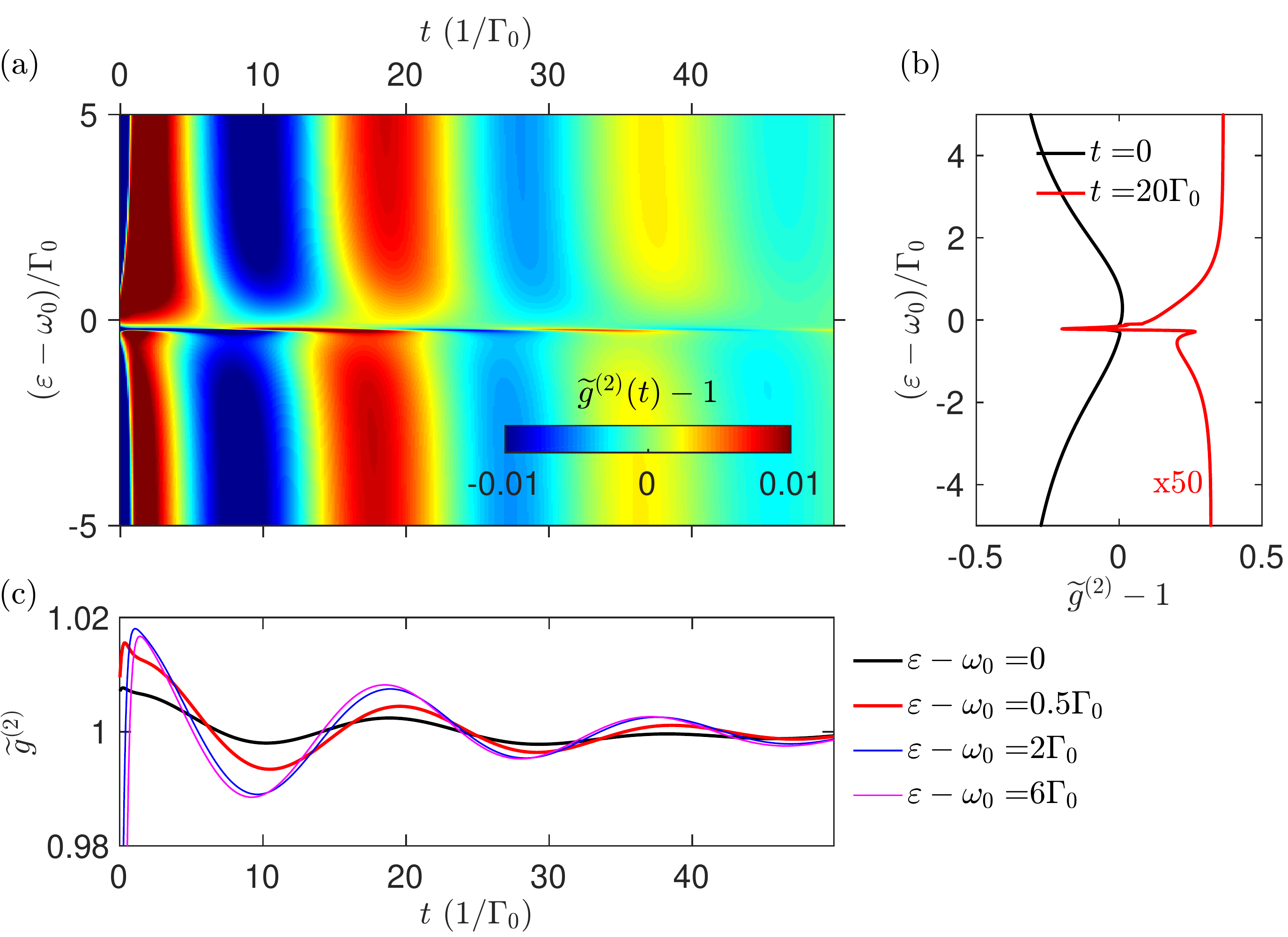}
	\caption{(a) Color map of photon-photon correlations Eq.~\eqref{eq:g2t} depending on time $t$ and average energy of two incident photons $\varepsilon$. (b) Spectra of photon-photon correlations for $t=0$ and $t=20\Gamma_{0}$. (c) Time dependence of photon-photon correlations for $\varepsilon-\omega_{0}=0, 0.5\Gamma_0, 2\Gamma_0, 6\Gamma_0$.
		Calculation has been performed for $(\omega_{1}-\omega_{2})/2\equiv \delta=20\Gamma_{0}$, $N=4$,
		$\varphi=\omega_{0}d/c=0.1$, and $\chi=10^4\Gamma_0$.
	}\label{fig:G2}
\end{figure}
%%%%%%%%%%%%
In this section we demonstrate how the superradiant, subradiant and twilight states are manifested in the time-dependent photon-photon correlations.
The wavefunction, describing the backscattering of the pair of photons, incident at the frequencies $\omega_{1}\ne\omega_{2}$, is given by
\begin{equation}
\psi_{2}=r_{1}^{\dag}r_{2}^{\dag}a_{-\omega_{1}}^{\dag}a_{-\omega_{2}}^{\dag}|0\rangle+\frac{\rmi}{2}\int\limits_{-\infty}^{\infty}
\frac{\rmd \omega}{2\pi}M(\varepsilon+\omega,\varepsilon-\omega,\omega_{1},\omega_{2})
a_{\varepsilon-\omega}^{\dag}a_{\varepsilon+\omega}^{\dag}|0\rangle\:.
\end{equation}
Here, the first term describes the coherent independent scattering of the photons, the second term accounts for the photon-photon correlations, and
$
r_{1,2}=-\rmi \Gamma_0 \sum_{i,j=1}^{N}G_{ij}(\varepsilon_{1,2})\e^{\rmi \omega_{0}/c(z_{i}+z_{j})}
$ are the reflection coefficients for individual photons.
We are interested in the time-dependent photon-photon correlations, that are given by
\begin{eqnarray}\label{eq:c}
c(t,\varepsilon)&&\equiv\langle \psi_{2}|a^{\dag}(0)a^{\dag}(t)a(t)a(0)|\psi_{2}\rangle
\nonumber \\
&&=\left|r_{1}r_{2}\cos\left(\frac{\omega_{1}-\omega_{2}}{2}t\right)+\frac{\rmi}{2}\int \frac{\rmd \omega}{2\pi}\e^{-\rmi \omega t}M(\varepsilon+\omega,\varepsilon-\omega,\omega_{1},\omega_{2})\right|^2\:,
\end{eqnarray}
where $a(t)=\int a_{\omega}\e^{-\i \omega t}d\omega$. We consider the situation when the total energy of the photon pair $2\varepsilon$
is varied around  $2\omega_{0}$, while the individual photon energies $\omega_{1,2}$ are strongly detuned from $\omega_{0}$. This allows us to  selectively and resonantly excite only the two-photon states.
Due to the strong detuning between $\omega_{1}$ and $\omega_{2}$, the first term in Eq.~\eqref{eq:c} will rapidly oscillate in time. Hence, we assume that the time-dependent correlations are
smoothed and defined in the following way:
\begin{equation}
\widetilde g^{(2)}(t,\varepsilon)=\frac{\mathcal Uc(t,\varepsilon)}{\lim_{t\to\infty}\mathcal U c(t,\varepsilon)},
\quad \mathcal U f(t)=f\left(t\to \frac{4\pi}{\omega_{1}-\omega_{2}}\left\lfloor \frac{\omega_{1}-\omega_{2}}{4\pi}t\right\rfloor\right )\:.
\end{equation}
The normalization has been explicitly chosen to satisfy the condition $\widetilde g^{(2)}(t\to\infty)=1$.
Since $\mathcal U \cos\left(\frac{\omega_{1}-\omega_{2}}{2}t\right)=1$, in the  regime when $|\omega_{1}-\omega_{2}|\gg\Gamma_{0}$ we obtain
\begin{equation}\label{eq:g2t}
\widetilde g^{(2)}(t,\varepsilon)=
\frac{\left| r_{1}r_{2}+\rmi\int \frac{\rmd \omega}{4\pi}\e^{-\rmi \omega t}M(\varepsilon+\omega,\varepsilon-\omega,\omega_{1},\omega_{2})\right|^{2}}{|r_{1}|^{2}|r_{2}|^{2}}\:.
\end{equation}
%%%%%%%%%%%%%%
We note, that $\widetilde g^{(2)}(t=0,\varepsilon)=0$ for $N=1$ due to the photon blockade effect~[\onlinecite{Poshakinskiy2016S}]. The scattering amplitude $M(\varepsilon+\omega,\varepsilon-\omega,\omega_{1},\omega_{2})$, defined in Eq.~\eqref{S}, depends on the energies of the scattered photons only via the single-particle Green functions,
$M\propto s_{i}(\varepsilon+\omega)s_{i}(\varepsilon-\omega)Q_{ij}(\varepsilon)$. As such, the lifetime of the correlations is determined only by the single-photon resonances. However,
the excitation efficiency of the different single-photon resonances still does depend on the photon pair energy $2\varepsilon$. This is the main ingredient of our proposal for observation of different two-photon states: when the pair energy $2\varepsilon$ is tuned to the double-excited subradiant or twilight state, the excitation efficiency of single-excited subradiant state increases, which results in long-lived photon-photon correlations. 

In order to test this approach, we have plotted in Fig.~\ref{fig:G2} the dependence of the $\widetilde g^{(2)}(t,\varepsilon)$ correlations on time $t$ and photon pair energy $2\varepsilon$. The calculation demonstrates,
that the spectra of the photon-photon correlations strongly depend on the delay time. Namely, at $t=0$ the spectrum
$\widetilde g^{(2)}(\varepsilon)$  is dominated by a broad feature with the half-width at half-maximum $\approx 3\Gamma_{0}$, corresponding to the excitation of the double-excited superradiant state [black curve in Fig.~\ref{fig:G2}(b)]. However, the superradiant mode  has short lifetime, and this broad feature is already vanished at $t\approx 1/\Gamma_{0}$. At larger time, $t=20/\Gamma_{0}$, the spectrum $\widetilde g^{(2)}(\varepsilon)$ becomes more narrow [red curve in Fig.~\ref{fig:G2}(b)]. The narrow features with the width $\sim \phi^{2}\Gamma_{0}\sim \Gamma_0/100\ll\Gamma_0$ correspond to the excitation of the two-particle subradiant states. The  wider features, with the width $\sim\Gamma_{0}$, are due to the twilight resonances.
%%%%%%%%%%
\begin{figure}[!htp]
	\centering\includegraphics[width=0.6\textwidth]{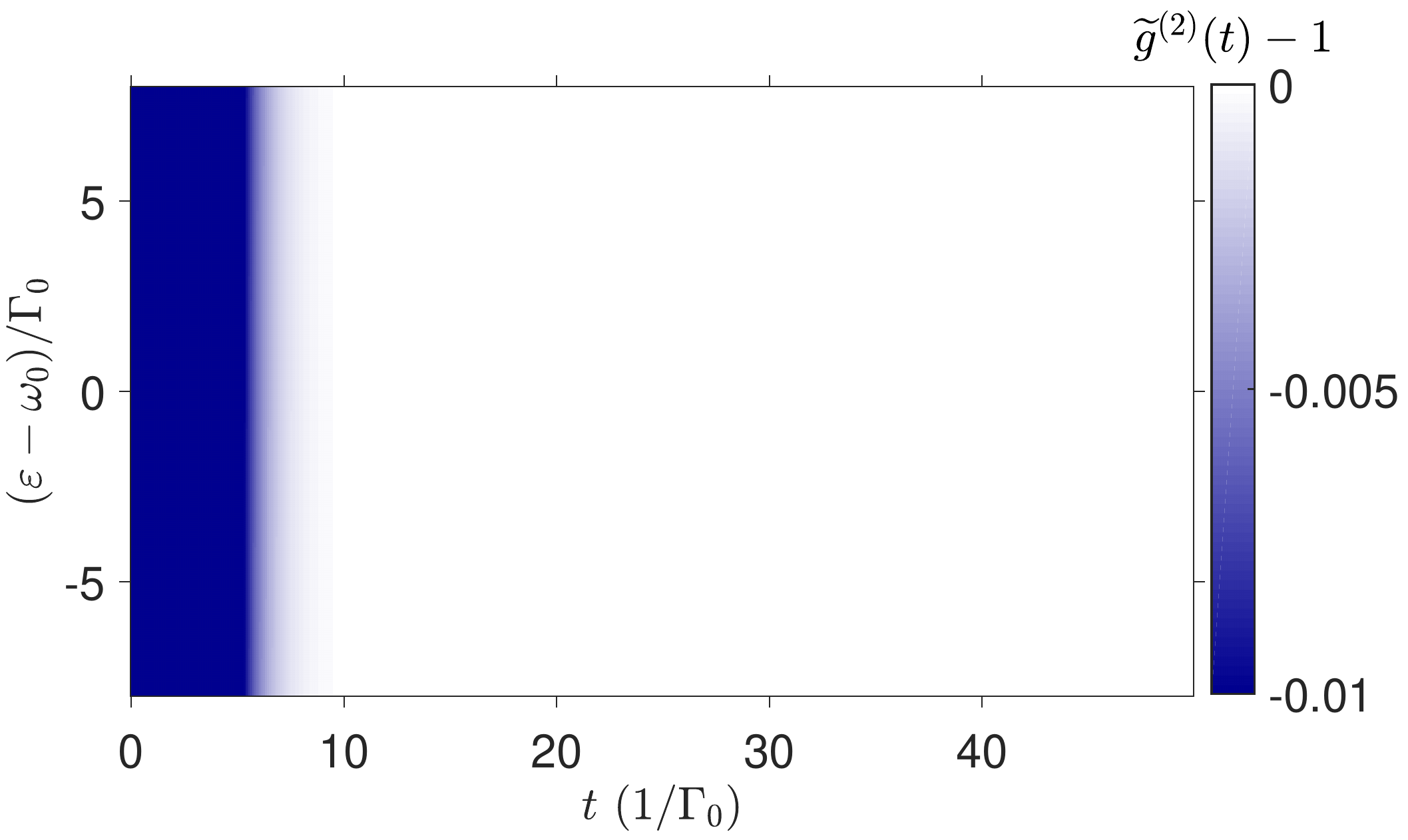}
	\caption{(a) Color map of photon-photon correlations as functions time $t$ and average energy of two incident photons $\varepsilon$ for a single qubit. 
		Calculation has been performed for $(\omega_{1}-\omega_{2})/2\equiv \delta=20\Gamma_{0}$, $N=1$, and $\chi=10^4\Gamma_0$.
	} \label{SingleG2}
\end{figure}
%%%%%%%%%%%%

The same contribution of the twilight states to the long-lived photon-photon correlations is also examined in Fig.~\ref{fig:G2}c that shows the curves  $\widetilde g^{(2)}(t,\varepsilon)$  as function of time for  different detunings $\eps-\omega_0$. For large detunings ($\eps-\omega_0=2\Gamma_0$ and $\eps-\omega_0=6\Gamma_0$, blue and magenta curves)  the correlations are practically independent of $\eps$. As such, these long-lived traces are not related to the two-photon states. They are due to  the subradiant single-particle resonances rather than double-excited states. However,  the amplitude of long-lived photon-photon correlations in  Fig.~\ref{fig:G2}(c)  is strongly modified when $\eps$ is tuned closer to $\omega_0$ (black and red curves). This is in full agreement with the results in Fig.~\ref{fig:G2}(b) and is explained by the excitation of the twilight states.

Thus, the twilight states can be used for excitation of long-lived photon-photon correlations. They show the same long lifetime as the two-particle  subradiant  states, $\sim1/(\phi^{2}\Gamma_{0})$ but are  relatively easier to excite due  to their broader spectral linewidth $\sim\Gamma_{0}$. The lifetime of photon-photon correlations is much longer than that of the individual qubits, see Fig.~\ref{SingleG2}. In the case of single qubit, the  photon-photon correlations quickly decay to $1$ with the decay rate $\Gamma_0$, and they are almost independent of the average energy of incident photons. This means that the waveguide photons coupled to qubits arrays enable more potential applications in storage and processing of quantum information.

\nocite{apsrev41Control}
\bibliographystyle{apsrev4}

%\bibliography{titleon,2phot}

\begin{thebibliography}{49}%
	\makeatletter
	\providecommand \@ifxundefined [1]{%
		\@ifx{#1\undefined}
	}%
	\providecommand \@ifnum [1]{%
		\ifnum #1\expandafter \@firstoftwo
		\else \expandafter \@secondoftwo
		\fi
	}%
	\providecommand \@ifx [1]{%
		\ifx #1\expandafter \@firstoftwo
		\else \expandafter \@secondoftwo
		\fi
	}%
	\providecommand \natexlab [1]{#1}%
	\providecommand \enquote  [1]{``#1''}%
	\providecommand \bibnamefont  [1]{#1}%
	\providecommand \bibfnamefont [1]{#1}%
	\providecommand \citenamefont [1]{#1}%
	\providecommand \href@noop [0]{\@secondoftwo}%
	\providecommand \href [0]{\begingroup \@sanitize@url \@href}%
	\providecommand \@href[1]{\@@startlink{#1}\@@href}%
	\providecommand \@@href[1]{\endgroup#1\@@endlink}%
	\providecommand \@sanitize@url [0]{\catcode `\\12\catcode `\$12\catcode
		`\&12\catcode `\#12\catcode `\^12\catcode `\_12\catcode `\%12\relax}%
	\providecommand \@@startlink[1]{}%
	\providecommand \@@endlink[0]{}%
	\providecommand \url  [0]{\begingroup\@sanitize@url \@url }%
	\providecommand \@url [1]{\endgroup\@href {#1}{\urlprefix }}%
	\providecommand \urlprefix  [0]{URL }%
	\providecommand \Eprint [0]{\href }%
	\providecommand \doibase [0]{http://dx.doi.org/}%
	\providecommand \selectlanguage [0]{\@gobble}%
	\providecommand \bibinfo  [0]{\@secondoftwo}%
	\providecommand \bibfield  [0]{\@secondoftwo}%
	\providecommand \translation [1]{[#1]}%
	\providecommand \BibitemOpen [0]{}%
	\providecommand \bibitemStop [0]{}%
	\providecommand \bibitemNoStop [0]{.\EOS\space}%
	\providecommand \EOS [0]{\spacefactor3000\relax}%
	\providecommand \BibitemShut  [1]{\csname bibitem#1\endcsname}%
	\let\auto@bib@innerbib\@empty
	%</preamble>
	\bibitem [{\citenamefont {Chang}\ \emph {et~al.}(2014)\citenamefont {Chang},
		\citenamefont {Vuleti{\'c}},\ and\ \citenamefont {Lukin}}]{chang2014}%
	\BibitemOpen
	\bibfield  {author} {\bibinfo {author} {\bibfnamefont {D.~E.}\ \bibnamefont
			{Chang}}, \bibinfo {author} {\bibfnamefont {V.}~\bibnamefont {Vuleti{\'c}}},
		\ and\ \bibinfo {author} {\bibfnamefont {M.~D.}\ \bibnamefont {Lukin}},\
	}\bibfield  {title} {\enquote {\bibinfo {title} {Quantum nonlinear optics ---
				photon by photon},}\ }\href
	{https://www.nature.com/articles/nphoton.2014.192} {\bibfield  {journal}
		{\bibinfo  {journal} {Nat. Photonics}\ }\textbf {\bibinfo {volume} {8}},\
		\bibinfo {pages} {685} (\bibinfo {year} {2014})}\BibitemShut {NoStop}%
	\bibitem [{\citenamefont {Dorfman}\ \emph {et~al.}(2016)\citenamefont
		{Dorfman}, \citenamefont {Schlawin},\ and\ \citenamefont
		{Mukamel}}]{Konstantin2016}%
	\BibitemOpen
	\bibfield  {author} {\bibinfo {author} {\bibfnamefont {K.~E.}\ \bibnamefont
			{Dorfman}}, \bibinfo {author} {\bibfnamefont {F.}~\bibnamefont {Schlawin}}, \
		and\ \bibinfo {author} {\bibfnamefont {S.}~\bibnamefont {Mukamel}},\
	}\bibfield  {title} {\enquote {\bibinfo {title} {Nonlinear optical signals
				and spectroscopy with quantum light},}\ }\href {\doibase
		10.1103/RevModPhys.88.045008} {\bibfield  {journal} {\bibinfo  {journal}
			{Rev. Mod. Phys.}\ }\textbf {\bibinfo {volume} {88}},\ \bibinfo {pages}
		{045008} (\bibinfo {year} {2016})}\BibitemShut {NoStop}%
	\bibitem [{\citenamefont {Kruk}\ \emph {et~al.}(2019)\citenamefont {Kruk},
		\citenamefont {Poddubny}, \citenamefont {Smirnova}, \citenamefont {Wang},
		\citenamefont {Slobozhanyuk}, \citenamefont {Shorokhov}, \citenamefont
		{Kravchenko}, \citenamefont {Luther-Davies},\ and\ \citenamefont
		{Kivshar}}]{Kruk2019}%
	\BibitemOpen
	\bibfield  {author} {\bibinfo {author} {\bibfnamefont {S.}~\bibnamefont
			{Kruk}}, \bibinfo {author} {\bibfnamefont {A.}~\bibnamefont {Poddubny}},
		\bibinfo {author} {\bibfnamefont {D.}~\bibnamefont {Smirnova}}, \bibinfo
		{author} {\bibfnamefont {L.}~\bibnamefont {Wang}}, \bibinfo {author}
		{\bibfnamefont {A.}~\bibnamefont {Slobozhanyuk}}, \bibinfo {author}
		{\bibfnamefont {A.}~\bibnamefont {Shorokhov}}, \bibinfo {author}
		{\bibfnamefont {I.}~\bibnamefont {Kravchenko}}, \bibinfo {author}
		{\bibfnamefont {B.}~\bibnamefont {Luther-Davies}}, \ and\ \bibinfo {author}
		{\bibfnamefont {Y.}~\bibnamefont {Kivshar}},\ }\bibfield  {title} {\enquote
		{\bibinfo {title} {Nonlinear light generation in topological
				nanostructures},}\ }\href {\doibase 10.1038/s41565-018-0324-7} {\bibfield
		{journal} {\bibinfo  {journal} {Nat. Nanotechnology}\ }\textbf {\bibinfo
			{volume} {14}},\ \bibinfo {pages} {126--130} (\bibinfo {year}
		{2019})}\BibitemShut {NoStop}%
	\bibitem [{\citenamefont {Kimble}(2008)}]{kimble2008quantum}%
	\BibitemOpen
	\bibfield  {author} {\bibinfo {author} {\bibfnamefont {H.~J.}\ \bibnamefont
			{Kimble}},\ }\bibfield  {title} {\enquote {\bibinfo {title} {The quantum
				internet},}\ }\href {https://www.nature.com/articles/nature07127} {\bibfield
		{journal} {\bibinfo  {journal} {Nature}\ }\textbf {\bibinfo {volume} {453}},\
		\bibinfo {pages} {1023} (\bibinfo {year} {2008})}\BibitemShut {NoStop}%
	\bibitem [{\citenamefont {Tittl}\ \emph {et~al.}(2018)\citenamefont {Tittl},
		\citenamefont {Leitis}, \citenamefont {Liu}, \citenamefont {Yesilkoy},
		\citenamefont {Choi}, \citenamefont {Neshev}, \citenamefont {Kivshar},\ and\
		\citenamefont {Altug}}]{Tittl2018}%
	\BibitemOpen
	\bibfield  {author} {\bibinfo {author} {\bibfnamefont {A.}~\bibnamefont
			{Tittl}}, \bibinfo {author} {\bibfnamefont {A.}~\bibnamefont {Leitis}},
		\bibinfo {author} {\bibfnamefont {M.}~\bibnamefont {Liu}}, \bibinfo {author}
		{\bibfnamefont {F.}~\bibnamefont {Yesilkoy}}, \bibinfo {author}
		{\bibfnamefont {D.-Y.}\ \bibnamefont {Choi}}, \bibinfo {author}
		{\bibfnamefont {D.~N.}\ \bibnamefont {Neshev}}, \bibinfo {author}
		{\bibfnamefont {Y.~S.}\ \bibnamefont {Kivshar}}, \ and\ \bibinfo {author}
		{\bibfnamefont {H.}~\bibnamefont {Altug}},\ }\bibfield  {title} {\enquote
		{\bibinfo {title} {Imaging-based molecular barcoding with pixelated
				dielectric metasurfaces},}\ }\href {\doibase 10.1126/science.aas9768}
	{\bibfield  {journal} {\bibinfo  {journal} {Science}\ }\textbf {\bibinfo
			{volume} {360}},\ \bibinfo {pages} {1105--1109} (\bibinfo {year}
		{2018})}\BibitemShut {NoStop}%
	\bibitem [{\citenamefont {Dicke}(1954)}]{Dicke1954}%
	\BibitemOpen
	\bibfield  {author} {\bibinfo {author} {\bibfnamefont {R.~H.}\ \bibnamefont
			{Dicke}},\ }\bibfield  {title} {\enquote {\bibinfo {title} {{C}oherence in
				{S}pontaneous {R}adiation {P}rocesses},}\ }\href {\doibase
		10.1103/PhysRev.93.99} {\bibfield  {journal} {\bibinfo  {journal} {Phys.
				Rev.}\ }\textbf {\bibinfo {volume} {93}},\ \bibinfo {pages} {99} (\bibinfo
		{year} {1954})}\BibitemShut {NoStop}%
	\bibitem [{\citenamefont {Roy}\ \emph {et~al.}(2017)\citenamefont {Roy},
		\citenamefont {Wilson},\ and\ \citenamefont {Firstenberg}}]{Roy2017}%
	\BibitemOpen
	\bibfield  {author} {\bibinfo {author} {\bibfnamefont {D.}~\bibnamefont
			{Roy}}, \bibinfo {author} {\bibfnamefont {C.~M.}\ \bibnamefont {Wilson}}, \
		and\ \bibinfo {author} {\bibfnamefont {O.}~\bibnamefont {Firstenberg}},\
	}\bibfield  {title} {\enquote {\bibinfo {title} {Colloquium: Strongly
				interacting photons in one-dimensional continuum},}\ }\href {\doibase
		10.1103/RevModPhys.89.021001} {\bibfield  {journal} {\bibinfo  {journal}
			{Rev. Mod. Phys.}\ }\textbf {\bibinfo {volume} {89}},\ \bibinfo {pages}
		{021001} (\bibinfo {year} {2017})}\BibitemShut {NoStop}%
	\bibitem [{\citenamefont {Chang}\ \emph {et~al.}(2018)\citenamefont {Chang},
		\citenamefont {Douglas}, \citenamefont {Gonz\'alez-Tudela}, \citenamefont
		{Hung},\ and\ \citenamefont {Kimble}}]{KimbleRMP2018}%
	\BibitemOpen
	\bibfield  {author} {\bibinfo {author} {\bibfnamefont {D.~E.}\ \bibnamefont
			{Chang}}, \bibinfo {author} {\bibfnamefont {J.~S.}\ \bibnamefont {Douglas}},
		\bibinfo {author} {\bibfnamefont {A.}~\bibnamefont {Gonz\'alez-Tudela}},
		\bibinfo {author} {\bibfnamefont {C.-L.}\ \bibnamefont {Hung}}, \ and\
		\bibinfo {author} {\bibfnamefont {H.~J.}\ \bibnamefont {Kimble}},\ }\bibfield
	{title} {\enquote {\bibinfo {title} {Colloquium: Quantum matter built from
				nanoscopic lattices of atoms and photons},}\ }\href {\doibase
		10.1103/RevModPhys.90.031002} {\bibfield  {journal} {\bibinfo  {journal}
			{Rev. Mod. Phys.}\ }\textbf {\bibinfo {volume} {90}},\ \bibinfo {pages}
		{031002} (\bibinfo {year} {2018})}\BibitemShut {NoStop}%
	\bibitem [{\citenamefont {Kockum}\ \emph {et~al.}(2019)\citenamefont {Kockum},
		\citenamefont {Miranowicz}, \citenamefont {Liberato}, \citenamefont
		{Savasta},\ and\ \citenamefont {Nori}}]{FriskKockum2019}%
	\BibitemOpen
	\bibfield  {author} {\bibinfo {author} {\bibfnamefont {A.~F.}\ \bibnamefont
			{Kockum}}, \bibinfo {author} {\bibfnamefont {A.}~\bibnamefont {Miranowicz}},
		\bibinfo {author} {\bibfnamefont {S.~D.}\ \bibnamefont {Liberato}}, \bibinfo
		{author} {\bibfnamefont {S.}~\bibnamefont {Savasta}}, \ and\ \bibinfo
		{author} {\bibfnamefont {F.}~\bibnamefont {Nori}},\ }\bibfield  {title}
	{\enquote {\bibinfo {title} {Ultrastrong coupling between light and
				matter},}\ }\href {\doibase 10.1038/s42254-018-0006-2} {\bibfield  {journal}
		{\bibinfo  {journal} {Nature Reviews Physics}\ }\textbf {\bibinfo {volume}
			{1}},\ \bibinfo {pages} {19--40} (\bibinfo {year} {2019})}\BibitemShut
	{NoStop}%
	\bibitem [{\citenamefont {{Ivchenko}}\ \emph {et~al.}(1994)\citenamefont
		{{Ivchenko}}, \citenamefont {{Nesvizhskii}},\ and\ \citenamefont
		{{Jorda}}}]{Ivchenko1994}%
	\BibitemOpen
	\bibfield  {author} {\bibinfo {author} {\bibfnamefont {E.~L.}\ \bibnamefont
			{{Ivchenko}}}, \bibinfo {author} {\bibfnamefont {A.~I.}\ \bibnamefont
			{{Nesvizhskii}}}, \ and\ \bibinfo {author} {\bibfnamefont {S.}~\bibnamefont
			{{Jorda}}},\ }\bibfield  {title} {\enquote {\bibinfo {title} {{B}ragg
				reflection of light from quantum-well structures},}\ }\href@noop {}
	{\bibfield  {journal} {\bibinfo  {journal} {Phys. Solid State}\ }\textbf
		{\bibinfo {volume} {36}},\ \bibinfo {pages} {1156--1161} (\bibinfo {year}
		{1994})}\BibitemShut {NoStop}%
	\bibitem [{\citenamefont {Birkl}\ \emph {et~al.}(1995)\citenamefont {Birkl},
		\citenamefont {Gatzke}, \citenamefont {Deutsch}, \citenamefont {Rolston},\
		and\ \citenamefont {Phillips}}]{Birkl1995}%
	\BibitemOpen
	\bibfield  {author} {\bibinfo {author} {\bibfnamefont {G.}~\bibnamefont
			{Birkl}}, \bibinfo {author} {\bibfnamefont {M.}~\bibnamefont {Gatzke}},
		\bibinfo {author} {\bibfnamefont {I.~H.}\ \bibnamefont {Deutsch}}, \bibinfo
		{author} {\bibfnamefont {S.~L.}\ \bibnamefont {Rolston}}, \ and\ \bibinfo
		{author} {\bibfnamefont {W.~D.}\ \bibnamefont {Phillips}},\ }\bibfield
	{title} {\enquote {\bibinfo {title} {Bragg scattering from atoms in optical
				lattices},}\ }\href {\doibase 10.1103/PhysRevLett.75.2823} {\bibfield
		{journal} {\bibinfo  {journal} {Phys. Rev. Lett.}\ }\textbf {\bibinfo
			{volume} {75}},\ \bibinfo {pages} {2823--2826} (\bibinfo {year}
		{1995})}\BibitemShut {NoStop}%
	\bibitem [{\citenamefont {DeVoe}\ and\ \citenamefont
		{Brewer}(1996)}]{Devoe1996}%
	\BibitemOpen
	\bibfield  {author} {\bibinfo {author} {\bibfnamefont {R.~G.}\ \bibnamefont
			{DeVoe}}\ and\ \bibinfo {author} {\bibfnamefont {R.~G.}\ \bibnamefont
			{Brewer}},\ }\bibfield  {title} {\enquote {\bibinfo {title} {Observation of
				superradiant and subradiant spontaneous emission of two trapped ions},}\
	}\href {\doibase 10.1103/PhysRevLett.76.2049} {\bibfield  {journal} {\bibinfo
			{journal} {Phys. Rev. Lett.}\ }\textbf {\bibinfo {volume} {76}},\ \bibinfo
		{pages} {2049--2052} (\bibinfo {year} {1996})}\BibitemShut {NoStop}%
	\bibitem [{\citenamefont {Chumakov}\ \emph {et~al.}(1999)\citenamefont
		{Chumakov}, \citenamefont {Niesen}, \citenamefont {Nagy},\ and\ \citenamefont
		{Alp}}]{chumakov1999}%
	\BibitemOpen
	\bibfield  {author} {\bibinfo {author} {\bibfnamefont {A.}~\bibnamefont
			{Chumakov}}, \bibinfo {author} {\bibfnamefont {L.}~\bibnamefont {Niesen}},
		\bibinfo {author} {\bibfnamefont {D.}~\bibnamefont {Nagy}}, \ and\ \bibinfo
		{author} {\bibfnamefont {E.}~\bibnamefont {Alp}},\ }\bibfield  {title}
	{\enquote {\bibinfo {title} {{N}uclear resonant scattering of synchrotron
				radiation by multilayer structures},}\ }\href@noop {} {\bibfield  {journal}
		{\bibinfo  {journal} {Hyperfine Interactions}\ }\textbf {\bibinfo {volume}
			{123-124}},\ \bibinfo {pages} {427--454} (\bibinfo {year}
		{1999})}\BibitemShut {NoStop}%
	\bibitem [{\citenamefont {Hendrickson}\ \emph {et~al.}(2008)\citenamefont
		{Hendrickson}, \citenamefont {Richards}, \citenamefont {Sweet}, \citenamefont
		{Khitrova}, \citenamefont {Poddubny}, \citenamefont {Ivchenko}, \citenamefont
		{Wegener},\ and\ \citenamefont {Gibbs}}]{Hendrickson2008}%
	\BibitemOpen
	\bibfield  {author} {\bibinfo {author} {\bibfnamefont {J.}~\bibnamefont
			{Hendrickson}}, \bibinfo {author} {\bibfnamefont {B.~C.}\ \bibnamefont
			{Richards}}, \bibinfo {author} {\bibfnamefont {J.}~\bibnamefont {Sweet}},
		\bibinfo {author} {\bibfnamefont {G.}~\bibnamefont {Khitrova}}, \bibinfo
		{author} {\bibfnamefont {A.~N.}\ \bibnamefont {Poddubny}}, \bibinfo {author}
		{\bibfnamefont {E.~L.}\ \bibnamefont {Ivchenko}}, \bibinfo {author}
		{\bibfnamefont {M.}~\bibnamefont {Wegener}}, \ and\ \bibinfo {author}
		{\bibfnamefont {H.~M.}\ \bibnamefont {Gibbs}},\ }\bibfield  {title} {\enquote
		{\bibinfo {title} {{E}xcitonic polaritons in $\mbox{F}$ibonacci
				quasicrystals},}\ }\href {\doibase 10.1364/OE.16.015382} {\bibfield
		{journal} {\bibinfo  {journal} {Opt. Express}\ }\textbf {\bibinfo {volume}
			{16}},\ \bibinfo {pages} {15382--15387} (\bibinfo {year} {2008})}\BibitemShut
	{NoStop}%
	\bibitem [{\citenamefont {{Goldberg}}\ \emph {et~al.}(2009)\citenamefont
		{{Goldberg}}, \citenamefont {{Deych}}, \citenamefont {{Lisyansky}},
		\citenamefont {{Shi}}, \citenamefont {{Menon}}, \citenamefont {{Tokranov}},
		\citenamefont {{Yakimov}},\ and\ \citenamefont
		{{Oktyabrsky}}}]{Goldberg2009}%
	\BibitemOpen
	\bibfield  {author} {\bibinfo {author} {\bibfnamefont {D.}~\bibnamefont
			{{Goldberg}}}, \bibinfo {author} {\bibfnamefont {L.~I.}\ \bibnamefont
			{{Deych}}}, \bibinfo {author} {\bibfnamefont {A.~A.}\ \bibnamefont
			{{Lisyansky}}}, \bibinfo {author} {\bibfnamefont {Z.}~\bibnamefont {{Shi}}},
		\bibinfo {author} {\bibfnamefont {V.~M.}\ \bibnamefont {{Menon}}}, \bibinfo
		{author} {\bibfnamefont {V.}~\bibnamefont {{Tokranov}}}, \bibinfo {author}
		{\bibfnamefont {M.}~\bibnamefont {{Yakimov}}}, \ and\ \bibinfo {author}
		{\bibfnamefont {S.}~\bibnamefont {{Oktyabrsky}}},\ }\bibfield  {title}
	{\enquote {\bibinfo {title} {{E}xciton-lattice polaritons in
				multiple-quantum-well-based photonic crystals},}\ }\href {\doibase
		10.1038/nphoton.2009.190} {\bibfield  {journal} {\bibinfo  {journal} {Nat.
				Photonics}\ }\textbf {\bibinfo {volume} {3}},\ \bibinfo {pages} {662--666}
		(\bibinfo {year} {2009})}\BibitemShut {NoStop}%
	\bibitem [{\citenamefont {van Loo}\ \emph {et~al.}(2013)\citenamefont {van
			Loo}, \citenamefont {Fedorov}, \citenamefont {Lalumiere}, \citenamefont
		{Sanders}, \citenamefont {Blais},\ and\ \citenamefont
		{Wallraff}}]{vanLoo2013}%
	\BibitemOpen
	\bibfield  {author} {\bibinfo {author} {\bibfnamefont {A.~F.}\ \bibnamefont
			{van Loo}}, \bibinfo {author} {\bibfnamefont {A.}~\bibnamefont {Fedorov}},
		\bibinfo {author} {\bibfnamefont {K.}~\bibnamefont {Lalumiere}}, \bibinfo
		{author} {\bibfnamefont {B.~C.}\ \bibnamefont {Sanders}}, \bibinfo {author}
		{\bibfnamefont {A.}~\bibnamefont {Blais}}, \ and\ \bibinfo {author}
		{\bibfnamefont {A.}~\bibnamefont {Wallraff}},\ }\bibfield  {title} {\enquote
		{\bibinfo {title} {Photon-mediated interactions between distant artificial
				atoms},}\ }\href {\doibase 10.1126/science.1244324} {\bibfield  {journal}
		{\bibinfo  {journal} {Science}\ }\textbf {\bibinfo {volume} {342}},\ \bibinfo
		{pages} {1494--1496} (\bibinfo {year} {2013})}\BibitemShut {NoStop}%
	\bibitem [{\citenamefont {Poddubny}\ and\ \citenamefont
		{Ivchenko}(2013)}]{Ivchenko2013}%
	\BibitemOpen
	\bibfield  {author} {\bibinfo {author} {\bibfnamefont {A.}~\bibnamefont
			{Poddubny}}\ and\ \bibinfo {author} {\bibfnamefont {E.}~\bibnamefont
			{Ivchenko}},\ }\bibfield  {title} {\enquote {\bibinfo {title} {{R}esonant
				diffraction of electromagnetic waves from solids (a review)},}\ }\href
	{\doibase 10.1134/S1063783413050120} {\bibfield  {journal} {\bibinfo
			{journal} {Phys. Solid State}\ }\textbf {\bibinfo {volume} {55}},\ \bibinfo
		{pages} {905--923} (\bibinfo {year} {2013})}\BibitemShut {NoStop}%
	\bibitem [{\citenamefont {Mlynek}\ \emph {et~al.}(2014)\citenamefont {Mlynek},
		\citenamefont {Abdumalikov}, \citenamefont {Eichler},\ and\ \citenamefont
		{Wallraff}}]{mlynek2014}%
	\BibitemOpen
	\bibfield  {author} {\bibinfo {author} {\bibfnamefont {J.~A.}\ \bibnamefont
			{Mlynek}}, \bibinfo {author} {\bibfnamefont {A.~A.}\ \bibnamefont
			{Abdumalikov}}, \bibinfo {author} {\bibfnamefont {C.}~\bibnamefont
			{Eichler}}, \ and\ \bibinfo {author} {\bibfnamefont {A.}~\bibnamefont
			{Wallraff}},\ }\bibfield  {title} {\enquote {\bibinfo {title} {Observation of
				{Dicke} superradiance for two artificial atoms in a cavity with high decay
				rate},}\ }\href {https://www.nature.com/articles/ncomms6186} {\bibfield
		{journal} {\bibinfo  {journal} {Nat. Comm.}\ }\textbf {\bibinfo {volume}
			{5}},\ \bibinfo {pages} {5186} (\bibinfo {year} {2014})}\BibitemShut
	{NoStop}%
	\bibitem [{\citenamefont {Guerin}\ \emph {et~al.}(2016)\citenamefont {Guerin},
		\citenamefont {Ara\'ujo},\ and\ \citenamefont {Kaiser}}]{Guerin2016}%
	\BibitemOpen
	\bibfield  {author} {\bibinfo {author} {\bibfnamefont {W.}~\bibnamefont
			{Guerin}}, \bibinfo {author} {\bibfnamefont {M.~O.}\ \bibnamefont
			{Ara\'ujo}}, \ and\ \bibinfo {author} {\bibfnamefont {R.}~\bibnamefont
			{Kaiser}},\ }\bibfield  {title} {\enquote {\bibinfo {title} {Subradiance in a
				large cloud of cold atoms},}\ }\href {\doibase
		10.1103/PhysRevLett.116.083601} {\bibfield  {journal} {\bibinfo  {journal}
			{Phys. Rev. Lett.}\ }\textbf {\bibinfo {volume} {116}},\ \bibinfo {pages}
		{083601} (\bibinfo {year} {2016})}\BibitemShut {NoStop}%
	\bibitem [{\citenamefont {Jenkins}\ \emph {et~al.}(2017)\citenamefont
		{Jenkins}, \citenamefont {Ruostekoski}, \citenamefont {Papasimakis},
		\citenamefont {Savo},\ and\ \citenamefont {Zheludev}}]{Jenkin2017}%
	\BibitemOpen
	\bibfield  {author} {\bibinfo {author} {\bibfnamefont {S.~D.}\ \bibnamefont
			{Jenkins}}, \bibinfo {author} {\bibfnamefont {J.}~\bibnamefont
			{Ruostekoski}}, \bibinfo {author} {\bibfnamefont {N.}~\bibnamefont
			{Papasimakis}}, \bibinfo {author} {\bibfnamefont {S.}~\bibnamefont {Savo}}, \
		and\ \bibinfo {author} {\bibfnamefont {N.~I.}\ \bibnamefont {Zheludev}},\
	}\bibfield  {title} {\enquote {\bibinfo {title} {Many-body subradiant
				excitations in metamaterial arrays: Experiment and theory},}\ }\href
	{\doibase 10.1103/PhysRevLett.119.053901} {\bibfield  {journal} {\bibinfo
			{journal} {Phys. Rev. Lett.}\ }\textbf {\bibinfo {volume} {119}},\ \bibinfo
		{pages} {053901} (\bibinfo {year} {2017})}\BibitemShut {NoStop}%
	\bibitem [{\citenamefont {Limonov}\ \emph {et~al.}(2017)\citenamefont
		{Limonov}, \citenamefont {Rybin}, \citenamefont {Poddubny},\ and\
		\citenamefont {Kivshar}}]{Limonov2017}%
	\BibitemOpen
	\bibfield  {author} {\bibinfo {author} {\bibfnamefont {M.~F.}\ \bibnamefont
			{Limonov}}, \bibinfo {author} {\bibfnamefont {M.~V.}\ \bibnamefont {Rybin}},
		\bibinfo {author} {\bibfnamefont {A.~N.}\ \bibnamefont {Poddubny}}, \ and\
		\bibinfo {author} {\bibfnamefont {Y.~S.}\ \bibnamefont {Kivshar}},\
	}\bibfield  {title} {\enquote {\bibinfo {title} {Fano resonances in
				photonics},}\ }\href {\doibase 10.1038/nphoton.2017.142} {\bibfield
		{journal} {\bibinfo  {journal} {Nat. Photonics}\ }\textbf {\bibinfo {volume}
			{11}},\ \bibinfo {pages} {543--554} (\bibinfo {year} {2017})}\BibitemShut
	{NoStop}%
	\bibitem [{\citenamefont {Wolf}\ \emph {et~al.}(2018)\citenamefont {Wolf},
		\citenamefont {Schuster}, \citenamefont {Schmidt}, \citenamefont {Slama},\
		and\ \citenamefont {Zimmermann}}]{Wolf2018}%
	\BibitemOpen
	\bibfield  {author} {\bibinfo {author} {\bibfnamefont {P.}~\bibnamefont
			{Wolf}}, \bibinfo {author} {\bibfnamefont {S.~C.}\ \bibnamefont {Schuster}},
		\bibinfo {author} {\bibfnamefont {D.}~\bibnamefont {Schmidt}}, \bibinfo
		{author} {\bibfnamefont {S.}~\bibnamefont {Slama}}, \ and\ \bibinfo {author}
		{\bibfnamefont {C.}~\bibnamefont {Zimmermann}},\ }\bibfield  {title}
	{\enquote {\bibinfo {title} {Observation of subradiant atomic momentum states
				with {Bose}-{Einstein} condensates in a recoil resolving optical ring
				resonator},}\ }\href {\doibase 10.1103/PhysRevLett.121.173602} {\bibfield
		{journal} {\bibinfo  {journal} {Phys. Rev. Lett.}\ }\textbf {\bibinfo
			{volume} {121}},\ \bibinfo {pages} {173602} (\bibinfo {year}
		{2018})}\BibitemShut {NoStop}%
	\bibitem [{\citenamefont {Weiss}\ \emph {et~al.}(2018)\citenamefont {Weiss},
		\citenamefont {Ara{\'{u}}jo}, \citenamefont {Kaiser},\ and\ \citenamefont
		{Guerin}}]{Weiss2018}%
	\BibitemOpen
	\bibfield  {author} {\bibinfo {author} {\bibfnamefont {P.}~\bibnamefont
			{Weiss}}, \bibinfo {author} {\bibfnamefont {M.~O.}\ \bibnamefont
			{Ara{\'{u}}jo}}, \bibinfo {author} {\bibfnamefont {R.}~\bibnamefont
			{Kaiser}}, \ and\ \bibinfo {author} {\bibfnamefont {W.}~\bibnamefont
			{Guerin}},\ }\bibfield  {title} {\enquote {\bibinfo {title} {Subradiance and
				radiation trapping in cold atoms},}\ }\href {\doibase
		10.1088/1367-2630/aac5d0} {\bibfield  {journal} {\bibinfo  {journal} {New J.
				Phys.}\ }\textbf {\bibinfo {volume} {20}},\ \bibinfo {pages} {063024}
		(\bibinfo {year} {2018})}\BibitemShut {NoStop}%
	\bibitem [{\citenamefont {{Wang}}\ \emph {et~al.}(2019)\citenamefont {{Wang}},
		\citenamefont {{Li}}, \citenamefont {{Feng}}, \citenamefont {{Song}},
		\citenamefont {{Song}}, \citenamefont {{Liu}}, \citenamefont {{Guo}},
		\citenamefont {{Zhang}}, \citenamefont {{Dong}}, \citenamefont {{Zheng}},
		\citenamefont {{Wang}},\ and\ \citenamefont {{Wang}}}]{Wang2019arxiv}%
	\BibitemOpen
	\bibfield  {author} {\bibinfo {author} {\bibfnamefont {Z.}~\bibnamefont
			{{Wang}}}, \bibinfo {author} {\bibfnamefont {H.}~\bibnamefont {{Li}}},
		\bibinfo {author} {\bibfnamefont {W.}~\bibnamefont {{Feng}}}, \bibinfo
		{author} {\bibfnamefont {X.}~\bibnamefont {{Song}}}, \bibinfo {author}
		{\bibfnamefont {C.}~\bibnamefont {{Song}}}, \bibinfo {author} {\bibfnamefont
			{W.}~\bibnamefont {{Liu}}}, \bibinfo {author} {\bibfnamefont
			{Q.}~\bibnamefont {{Guo}}}, \bibinfo {author} {\bibfnamefont
			{X.}~\bibnamefont {{Zhang}}}, \bibinfo {author} {\bibfnamefont
			{H.}~\bibnamefont {{Dong}}}, \bibinfo {author} {\bibfnamefont
			{D.}~\bibnamefont {{Zheng}}}, \bibinfo {author} {\bibfnamefont
			{H.}~\bibnamefont {{Wang}}}, \ and\ \bibinfo {author} {\bibfnamefont {D.-W.}\
			\bibnamefont {{Wang}}},\ }\bibfield  {title} {\enquote {\bibinfo {title}
			{{Generation and controllable switching of superradiant and subradiant states
					in a 10-qubit superconducting circuit}},}\ }\href@noop {} {\bibfield
		{journal} {\bibinfo  {journal} {arXiv e-prints}\ ,\ \bibinfo {eid}
			{arXiv:1907.13468}} (\bibinfo {year} {2019})},\ \Eprint
	{http://arxiv.org/abs/1907.13468} {arXiv:1907.13468 [quant-ph]} \BibitemShut
	{NoStop}%
	\bibitem [{\citenamefont {Carletti}\ \emph {et~al.}(2018)\citenamefont
		{Carletti}, \citenamefont {Koshelev}, \citenamefont {De~Angelis},\ and\
		\citenamefont {Kivshar}}]{Carletti2018}%
	\BibitemOpen
	\bibfield  {author} {\bibinfo {author} {\bibfnamefont {L.}~\bibnamefont
			{Carletti}}, \bibinfo {author} {\bibfnamefont {K.}~\bibnamefont {Koshelev}},
		\bibinfo {author} {\bibfnamefont {C.}~\bibnamefont {De~Angelis}}, \ and\
		\bibinfo {author} {\bibfnamefont {Y.}~\bibnamefont {Kivshar}},\ }\bibfield
	{title} {\enquote {\bibinfo {title} {Giant nonlinear response at the
				nanoscale driven by bound states in the continuum},}\ }\href {\doibase
		10.1103/PhysRevLett.121.033903} {\bibfield  {journal} {\bibinfo  {journal}
			{Phys. Rev. Lett.}\ }\textbf {\bibinfo {volume} {121}},\ \bibinfo {pages}
		{033903} (\bibinfo {year} {2018})}\BibitemShut {NoStop}%
	\bibitem [{\citenamefont {{Poddubny}}\ and\ \citenamefont
		{{Smirnova}}(2018)}]{ANP2018}%
	\BibitemOpen
	\bibfield  {author} {\bibinfo {author} {\bibfnamefont {A.~N.}\ \bibnamefont
			{{Poddubny}}}\ and\ \bibinfo {author} {\bibfnamefont {D.~A.}\ \bibnamefont
			{{Smirnova}}},\ }\bibfield  {title} {\enquote {\bibinfo {title} {{Nonlinear
					generation of quantum-entangled photons from high-Q states in dielectric
					nanoparticles}},}\ }\href
	{https://ui.adsabs.harvard.edu/abs/2018arXiv180804811P} {\bibfield  {journal}
		{\bibinfo  {journal} {arXiv:1808.04811}\ } (\bibinfo {year}
		{2018})}\BibitemShut {NoStop}%
	\bibitem [{\citenamefont {Koshelev}\ \emph {et~al.}(2019)\citenamefont
		{Koshelev}, \citenamefont {Kruk}, \citenamefont {Choi}, \citenamefont
		{Melik-Gaykazyan}, \citenamefont {Smirnova}, \citenamefont {Park},\ and\
		\citenamefont {Kivshar}}]{Koshelev2019}%
	\BibitemOpen
	\bibfield  {author} {\bibinfo {author} {\bibfnamefont {K.}~\bibnamefont
			{Koshelev}}, \bibinfo {author} {\bibfnamefont {S.}~\bibnamefont {Kruk}},
		\bibinfo {author} {\bibfnamefont {J.-H.}\ \bibnamefont {Choi}}, \bibinfo
		{author} {\bibfnamefont {E.~V.}\ \bibnamefont {Melik-Gaykazyan}}, \bibinfo
		{author} {\bibfnamefont {D.}~\bibnamefont {Smirnova}}, \bibinfo {author}
		{\bibfnamefont {H.-G.}\ \bibnamefont {Park}}, \ and\ \bibinfo {author}
		{\bibfnamefont {Y.}~\bibnamefont {Kivshar}},\ }\bibfield  {title} {\enquote
		{\bibinfo {title} {Observation of extraordinary {SHG} from all-dielectric
				nanoantennas governed by bound states in the continuum},}\ }in\ \href
	{\doibase 10.1364/CLEO_QELS.2019.FW4B.3} {\emph {\bibinfo {booktitle} {Conf.
				Lasers Electro-Optics}}}\ (\bibinfo  {publisher} {Optical Society of
		America},\ \bibinfo {year} {2019})\ p.\ \bibinfo {pages} {FW4B.3}\BibitemShut
	{NoStop}%
	\bibitem [{\citenamefont {Yudson}\ and\ \citenamefont
		{Rupasov}(1984)}]{Yudson1984}%
	\BibitemOpen
	\bibfield  {author} {\bibinfo {author} {\bibfnamefont {V.}~\bibnamefont
			{Yudson}}\ and\ \bibinfo {author} {\bibfnamefont {V.}~\bibnamefont
			{Rupasov}},\ }\bibfield  {title} {\enquote {\bibinfo {title} {Exact {Dicke}
				superradiance theory: Bethe wavefunctions in the discrete atom model},}\
	}\href@noop {} {\bibfield  {journal} {\bibinfo  {journal} {Sov. Phys. JETP}\
		}\textbf {\bibinfo {volume} {59}},\ \bibinfo {pages} {478} (\bibinfo {year}
		{1984})}\BibitemShut {NoStop}%
	\bibitem [{\citenamefont {Shen}\ and\ \citenamefont
		{Fan}(2007{\natexlab{a}})}]{Shen2007}%
	\BibitemOpen
	\bibfield  {author} {\bibinfo {author} {\bibfnamefont {J.-T.}\ \bibnamefont
			{Shen}}\ and\ \bibinfo {author} {\bibfnamefont {S.}~\bibnamefont {Fan}},\
	}\bibfield  {title} {\enquote {\bibinfo {title} {Strongly correlated
				multiparticle transport in one dimension through a quantum impurity},}\
	}\href {\doibase 10.1103/PhysRevA.76.062709} {\bibfield  {journal} {\bibinfo
			{journal} {Phys. Rev. A}\ }\textbf {\bibinfo {volume} {76}},\ \bibinfo
		{pages} {062709} (\bibinfo {year} {2007}{\natexlab{a}})}\BibitemShut
	{NoStop}%
	\bibitem [{\citenamefont {Shen}\ and\ \citenamefont
		{Fan}(2007{\natexlab{b}})}]{Fan2007}%
	\BibitemOpen
	\bibfield  {author} {\bibinfo {author} {\bibfnamefont {J.-T.}\ \bibnamefont
			{Shen}}\ and\ \bibinfo {author} {\bibfnamefont {S.}~\bibnamefont {Fan}},\
	}\bibfield  {title} {\enquote {\bibinfo {title} {Strongly correlated
				two-photon transport in a one-dimensional waveguide coupled to a two-level
				system},}\ }\href {\doibase 10.1103/PhysRevLett.98.153003} {\bibfield
		{journal} {\bibinfo  {journal} {Phys. Rev. Lett.}\ }\textbf {\bibinfo
			{volume} {98}},\ \bibinfo {pages} {153003} (\bibinfo {year}
		{2007}{\natexlab{b}})}\BibitemShut {NoStop}%
	\bibitem [{\citenamefont {Fang}\ and\ \citenamefont
		{Baranger}(2015)}]{Baranger2015}%
	\BibitemOpen
	\bibfield  {author} {\bibinfo {author} {\bibfnamefont {Y.-L.~L.}\
			\bibnamefont {Fang}}\ and\ \bibinfo {author} {\bibfnamefont {H.~U.}\
			\bibnamefont {Baranger}},\ }\bibfield  {title} {\enquote {\bibinfo {title}
			{Waveguide {QED}: {Power} spectra and correlations of two photons scattered
				off multiple distant qubits and a mirror},}\ }\href {\doibase
		10.1103/PhysRevA.91.053845} {\bibfield  {journal} {\bibinfo  {journal} {Phys.
				Rev. A}\ }\textbf {\bibinfo {volume} {91}},\ \bibinfo {pages} {053845}
		(\bibinfo {year} {2015})}\BibitemShut {NoStop}%
	\bibitem [{\citenamefont {Albrecht}\ \emph {et~al.}(2019)\citenamefont
		{Albrecht}, \citenamefont {Henriet}, \citenamefont {Asenjo-Garcia},
		\citenamefont {Dieterle}, \citenamefont {Painter},\ and\ \citenamefont
		{Chang}}]{Albrecht2019}%
	\BibitemOpen
	\bibfield  {author} {\bibinfo {author} {\bibfnamefont {A.}~\bibnamefont
			{Albrecht}}, \bibinfo {author} {\bibfnamefont {L.}~\bibnamefont {Henriet}},
		\bibinfo {author} {\bibfnamefont {A.}~\bibnamefont {Asenjo-Garcia}}, \bibinfo
		{author} {\bibfnamefont {P.~B.}\ \bibnamefont {Dieterle}}, \bibinfo {author}
		{\bibfnamefont {O.}~\bibnamefont {Painter}}, \ and\ \bibinfo {author}
		{\bibfnamefont {D.~E.}\ \bibnamefont {Chang}},\ }\bibfield  {title} {\enquote
		{\bibinfo {title} {Subradiant states of quantum bits coupled to a
				one-dimensional waveguide},}\ }\href {\doibase 10.1088/1367-2630/ab0134}
	{\bibfield  {journal} {\bibinfo  {journal} {New J. Phys.}\ }\textbf {\bibinfo
			{volume} {21}},\ \bibinfo {pages} {025003} (\bibinfo {year}
		{2019})}\BibitemShut {NoStop}%
	\bibitem [{\citenamefont {Calaj\'o}\ \emph {et~al.}(2019)\citenamefont
		{Calaj\'o}, \citenamefont {Fang}, \citenamefont {Baranger},\ and\
		\citenamefont {Ciccarello}}]{Baranger2019b}%
	\BibitemOpen
	\bibfield  {author} {\bibinfo {author} {\bibfnamefont {G.}~\bibnamefont
			{Calaj\'o}}, \bibinfo {author} {\bibfnamefont {Y.-L.~L.}\ \bibnamefont
			{Fang}}, \bibinfo {author} {\bibfnamefont {H.~U.}\ \bibnamefont {Baranger}},
		\ and\ \bibinfo {author} {\bibfnamefont {F.}~\bibnamefont {Ciccarello}},\
	}\bibfield  {title} {\enquote {\bibinfo {title} {Exciting a bound state in
				the continuum through multiphoton scattering plus delayed quantum
				feedback},}\ }\href {\doibase 10.1103/PhysRevLett.122.073601} {\bibfield
		{journal} {\bibinfo  {journal} {Phys. Rev. Lett.}\ }\textbf {\bibinfo
			{volume} {122}},\ \bibinfo {pages} {073601} (\bibinfo {year}
		{2019})}\BibitemShut {NoStop}%
	\bibitem [{\citenamefont {Zhang}\ and\ \citenamefont
		{M\o{}lmer}(2019)}]{Molmer2019}%
	\BibitemOpen
	\bibfield  {author} {\bibinfo {author} {\bibfnamefont {Y.-X.}\ \bibnamefont
			{Zhang}}\ and\ \bibinfo {author} {\bibfnamefont {K.}~\bibnamefont
			{M\o{}lmer}},\ }\bibfield  {title} {\enquote {\bibinfo {title} {Theory of
				subradiant states of a one-dimensional two-level atom chain},}\ }\href
	{\doibase 10.1103/PhysRevLett.122.203605} {\bibfield  {journal} {\bibinfo
			{journal} {Phys. Rev. Lett.}\ }\textbf {\bibinfo {volume} {122}},\ \bibinfo
		{pages} {203605} (\bibinfo {year} {2019})}\BibitemShut {NoStop}%
	\bibitem [{\citenamefont {Henriet}\ \emph {et~al.}(2019)\citenamefont
		{Henriet}, \citenamefont {Douglas}, \citenamefont {Chang},\ and\
		\citenamefont {Albrecht}}]{Chang2019}%
	\BibitemOpen
	\bibfield  {author} {\bibinfo {author} {\bibfnamefont {L.}~\bibnamefont
			{Henriet}}, \bibinfo {author} {\bibfnamefont {J.~S.}\ \bibnamefont
			{Douglas}}, \bibinfo {author} {\bibfnamefont {D.~E.}\ \bibnamefont {Chang}},
		\ and\ \bibinfo {author} {\bibfnamefont {A.}~\bibnamefont {Albrecht}},\
	}\bibfield  {title} {\enquote {\bibinfo {title} {Critical open-system
				dynamics in a one-dimensional optical-lattice clock},}\ }\href {\doibase
		10.1103/PhysRevA.99.023802} {\bibfield  {journal} {\bibinfo  {journal} {Phys.
				Rev. A}\ }\textbf {\bibinfo {volume} {99}},\ \bibinfo {pages} {023802}
		(\bibinfo {year} {2019})}\BibitemShut {NoStop}%
	\bibitem [{\citenamefont {Zhang}\ \emph {et~al.}(2019)\citenamefont {Zhang},
		\citenamefont {Yu},\ and\ \citenamefont {M{\o}lmer}}]{zhang2019subradiant}%
	\BibitemOpen
	\bibfield  {author} {\bibinfo {author} {\bibfnamefont {Y.-X.}\ \bibnamefont
			{Zhang}}, \bibinfo {author} {\bibfnamefont {C.}~\bibnamefont {Yu}}, \ and\
		\bibinfo {author} {\bibfnamefont {K.}~\bibnamefont {M{\o}lmer}},\ }\bibfield
	{title} {\enquote {\bibinfo {title} {Subradiant dimer excited states of atom
				chains coupled to a {1D} waveguide},}\ }\href
	{https://arxiv.org/abs/1908.01818} {\bibfield  {journal} {\bibinfo  {journal}
			{arXiv:1908.01818}\ } (\bibinfo {year} {2019})}\BibitemShut {NoStop}%
	\bibitem [{\citenamefont {Yudson}\ and\ \citenamefont
		{Reineker}(2008)}]{Yudson2008}%
	\BibitemOpen
	\bibfield  {author} {\bibinfo {author} {\bibfnamefont {V.~I.}\ \bibnamefont
			{Yudson}}\ and\ \bibinfo {author} {\bibfnamefont {P.}~\bibnamefont
			{Reineker}},\ }\bibfield  {title} {\enquote {\bibinfo {title} {Multiphoton
				scattering in a one-dimensional waveguide with resonant atoms},}\ }\href
	{\doibase 10.1103/PhysRevA.78.052713} {\bibfield  {journal} {\bibinfo
			{journal} {Phys. Rev. A}\ }\textbf {\bibinfo {volume} {78}},\ \bibinfo
		{pages} {052713} (\bibinfo {year} {2008})}\BibitemShut {NoStop}%
	\bibitem [{\citenamefont {Firstenberg}\ \emph {et~al.}(2013)\citenamefont
		{Firstenberg}, \citenamefont {Peyronel}, \citenamefont {Liang}, \citenamefont
		{Gorshkov}, \citenamefont {Lukin},\ and\ \citenamefont
		{Vuleti{\'{c}}}}]{Firstenberg2013}%
	\BibitemOpen
	\bibfield  {author} {\bibinfo {author} {\bibfnamefont {O.}~\bibnamefont
			{Firstenberg}}, \bibinfo {author} {\bibfnamefont {T.}~\bibnamefont
			{Peyronel}}, \bibinfo {author} {\bibfnamefont {Q.-Y.}\ \bibnamefont {Liang}},
		\bibinfo {author} {\bibfnamefont {A.~V.}\ \bibnamefont {Gorshkov}}, \bibinfo
		{author} {\bibfnamefont {M.~D.}\ \bibnamefont {Lukin}}, \ and\ \bibinfo
		{author} {\bibfnamefont {V.}~\bibnamefont {Vuleti{\'{c}}}},\ }\bibfield
	{title} {\enquote {\bibinfo {title} {Attractive photons in a quantum
				nonlinear medium},}\ }\href {\doibase 10.1038/nature12512} {\bibfield
		{journal} {\bibinfo  {journal} {Nature}\ }\textbf {\bibinfo {volume} {502}},\
		\bibinfo {pages} {71--75} (\bibinfo {year} {2013})}\BibitemShut {NoStop}%
	\bibitem [{\citenamefont {Laakso}\ and\ \citenamefont
		{Pletyukhov}(2014)}]{Laakso2014}%
	\BibitemOpen
	\bibfield  {author} {\bibinfo {author} {\bibfnamefont {M.}~\bibnamefont
			{Laakso}}\ and\ \bibinfo {author} {\bibfnamefont {M.}~\bibnamefont
			{Pletyukhov}},\ }\bibfield  {title} {\enquote {\bibinfo {title} {Scattering
				of two photons from two distant qubits: Exact solution},}\ }\href {\doibase
		10.1103/PhysRevLett.113.183601} {\bibfield  {journal} {\bibinfo  {journal}
			{Phys. Rev. Lett.}\ }\textbf {\bibinfo {volume} {113}},\ \bibinfo {pages}
		{183601} (\bibinfo {year} {2014})}\BibitemShut {NoStop}%
	\bibitem [{\citenamefont {Fang}\ \emph {et~al.}(2014)\citenamefont {Fang},
		\citenamefont {Zheng},\ and\ \citenamefont {Baranger}}]{Fang2014}%
	\BibitemOpen
	\bibfield  {author} {\bibinfo {author} {\bibfnamefont {Y.-L.~L.}\
			\bibnamefont {Fang}}, \bibinfo {author} {\bibfnamefont {H.}~\bibnamefont
			{Zheng}}, \ and\ \bibinfo {author} {\bibfnamefont {H.~U.}\ \bibnamefont
			{Baranger}},\ }\bibfield  {title} {\enquote {\bibinfo {title}
			{One-dimensional waveguide coupled to multiple qubits: photon-photon
				correlations},}\ }\href {https://doi.org/10.1140/epjqt3} {\bibfield
		{journal} {\bibinfo  {journal} {{EPJ} Quantum Technology}\ }\textbf {\bibinfo
			{volume} {1}},\ \bibinfo {pages} {3} (\bibinfo {year} {2014})}\BibitemShut
	{NoStop}%
	\bibitem [{\citenamefont {Xu}\ and\ \citenamefont {Fan}(2015)}]{Xu2015}%
	\BibitemOpen
	\bibfield  {author} {\bibinfo {author} {\bibfnamefont {S.}~\bibnamefont
			{Xu}}\ and\ \bibinfo {author} {\bibfnamefont {S.}~\bibnamefont {Fan}},\
	}\bibfield  {title} {\enquote {\bibinfo {title} {Input-output formalism for
				few-photon transport: A systematic treatment beyond two photons},}\ }\href
	{\doibase 10.1103/PhysRevA.91.043845} {\bibfield  {journal} {\bibinfo
			{journal} {Phys. Rev. A}\ }\textbf {\bibinfo {volume} {91}},\ \bibinfo
		{pages} {043845} (\bibinfo {year} {2015})}\BibitemShut {NoStop}%
	\bibitem [{\citenamefont {Muthukrishnan}\ \emph {et~al.}(2004)\citenamefont
		{Muthukrishnan}, \citenamefont {Agarwal},\ and\ \citenamefont
		{Scully}}]{Muthukrishnan2004}%
	\BibitemOpen
	\bibfield  {author} {\bibinfo {author} {\bibfnamefont {A.}~\bibnamefont
			{Muthukrishnan}}, \bibinfo {author} {\bibfnamefont {G.~S.}\ \bibnamefont
			{Agarwal}}, \ and\ \bibinfo {author} {\bibfnamefont {M.~O.}\ \bibnamefont
			{Scully}},\ }\bibfield  {title} {\enquote {\bibinfo {title} {Inducing
				disallowed two-atom transitions with temporally entangled photons},}\ }\href
	{\doibase 10.1103/PhysRevLett.93.093002} {\bibfield  {journal} {\bibinfo
			{journal} {Phys. Rev. Lett.}\ }\textbf {\bibinfo {volume} {93}},\ \bibinfo
		{pages} {093002} (\bibinfo {year} {2004})}\BibitemShut {NoStop}%
	\bibitem [{\citenamefont {Zheng}\ and\ \citenamefont
		{Baranger}(2013)}]{Baranger2013}%
	\BibitemOpen
	\bibfield  {author} {\bibinfo {author} {\bibfnamefont {H.}~\bibnamefont
			{Zheng}}\ and\ \bibinfo {author} {\bibfnamefont {H.~U.}\ \bibnamefont
			{Baranger}},\ }\bibfield  {title} {\enquote {\bibinfo {title} {Persistent
				quantum beats and long-distance entanglement from waveguide-mediated
				interactions},}\ }\href {\doibase 10.1103/PhysRevLett.110.113601} {\bibfield
		{journal} {\bibinfo  {journal} {Phys. Rev. Lett.}\ }\textbf {\bibinfo
			{volume} {110}},\ \bibinfo {pages} {113601} (\bibinfo {year}
		{2013})}\BibitemShut {NoStop}%
	\bibitem [{\citenamefont {Poshakinskiy}\ and\ \citenamefont
		{Poddubny}(2016)}]{Poshakinskiy2016}%
	\BibitemOpen
	\bibfield  {author} {\bibinfo {author} {\bibfnamefont {A.~V.}\ \bibnamefont
			{Poshakinskiy}}\ and\ \bibinfo {author} {\bibfnamefont {A.~N.}\ \bibnamefont
			{Poddubny}},\ }\bibfield  {title} {\enquote {\bibinfo {title}
			{Biexciton-mediated superradiant photon blockade},}\ }\href {\doibase
		10.1103/PhysRevA.93.033856} {\bibfield  {journal} {\bibinfo  {journal} {Phys.
				Rev. A}\ }\textbf {\bibinfo {volume} {93}},\ \bibinfo {pages} {033856}
		(\bibinfo {year} {2016})}\BibitemShut {NoStop}%
	\bibitem [{Sup()}]{Suppl}%
	\BibitemOpen
	\href@noop {} {\bibinfo  {journal} {See Supplemental Material for details of
			(S1) Effective model for the excitations; (S2) Decay process; (S3) Subradiant
			states; (S4) Two-photon scattering; (S5) Threshold of qubit number for
			subradiant states; (S6) Subradiant and twilight states in photon-photon
			correlations, which includes Refs.~[32], [34], [36], [44], [46] and [47]}\ }\BibitemShut
	{NoStop}%
	\bibitem [{\citenamefont {Kashcheyevs}\ and\ \citenamefont
		{Kaestner}(2010)}]{Kashcheyevs2016}%
	\BibitemOpen
	\bibfield  {author} {\bibinfo {author} {\bibfnamefont {V.}~\bibnamefont
			{Kashcheyevs}}\ and\ \bibinfo {author} {\bibfnamefont {B.}~\bibnamefont
			{Kaestner}},\ }\bibfield  {title} {\enquote {\bibinfo {title} {Universal
				decay cascade model for dynamic quantum dot initialization},}\ }\href
	{\doibase 10.1103/PhysRevLett.104.186805} {\bibfield  {journal} {\bibinfo
			{journal} {Phys. Rev. Lett.}\ }\textbf {\bibinfo {volume} {104}},\ \bibinfo
		{pages} {186805} (\bibinfo {year} {2010})}\BibitemShut {NoStop}%
	\bibitem [{\citenamefont {Ivchenko}(2005)}]{Ivchenko2005}%
	\BibitemOpen
	\bibfield  {journal} {  }\bibfield  {author} {\bibinfo {author} {\bibfnamefont
			{E.~L.}\ \bibnamefont {Ivchenko}},\ }\href@noop {} {\emph {\bibinfo {title}
			{{O}ptical {S}pectroscopy of {S}emiconductor {N}anostructures}}}\ (\bibinfo
	{publisher} {Alpha Science International},\ \bibinfo {address} {Harrow, UK},\
	\bibinfo {year} {2005})\BibitemShut {NoStop}%
	\bibitem [{\citenamefont {Kocaba\ifmmode~\mbox{\c{s}}\else
			\c{s}\fi{}}(2016)}]{Kocabas2016}%
	\BibitemOpen
	\bibfield  {author} {\bibinfo {author} {\bibfnamefont {{\ifmmode
					\mbox{\c{S}}\else \c{S}\fi{}}.~E.}\ \bibnamefont
			{Kocaba\ifmmode~\mbox{\c{s}}\else \c{s}\fi{}}},\ }\bibfield  {title}
	{\enquote {\bibinfo {title} {Effects of modal dispersion on few-photon--qubit
				scattering in one-dimensional waveguides},}\ }\href {\doibase
		10.1103/PhysRevA.93.033829} {\bibfield  {journal} {\bibinfo  {journal} {Phys.
				Rev. A}\ }\textbf {\bibinfo {volume} {93}},\ \bibinfo {pages} {033829}
		(\bibinfo {year} {2016})}\BibitemShut {NoStop}%
	\bibitem [{\citenamefont {Schneider}\ \emph {et~al.}(2016)\citenamefont
		{Schneider}, \citenamefont {Sproll}, \citenamefont {Stawiarski},
		\citenamefont {Schmitteckert},\ and\ \citenamefont {Busch}}]{Schneider2016}%
	\BibitemOpen
	\bibfield  {author} {\bibinfo {author} {\bibfnamefont {M.~P.}\ \bibnamefont
			{Schneider}}, \bibinfo {author} {\bibfnamefont {T.}~\bibnamefont {Sproll}},
		\bibinfo {author} {\bibfnamefont {C.}~\bibnamefont {Stawiarski}}, \bibinfo
		{author} {\bibfnamefont {P.}~\bibnamefont {Schmitteckert}}, \ and\ \bibinfo
		{author} {\bibfnamefont {K.}~\bibnamefont {Busch}},\ }\bibfield  {title}
	{\enquote {\bibinfo {title} {Green's-function formalism for waveguide {QED}
				applications},}\ }\href {\doibase 10.1103/PhysRevA.93.013828} {\bibfield
		{journal} {\bibinfo  {journal} {Phys. Rev. A}\ }\textbf {\bibinfo {volume}
			{93}},\ \bibinfo {pages} {013828} (\bibinfo {year} {2016})}\BibitemShut
	{NoStop}%
	\bibitem [{\citenamefont {Liang}\ \emph {et~al.}(2018)\citenamefont {Liang},
		\citenamefont {Venkatramani}, \citenamefont {Cantu}, \citenamefont
		{Nicholson}, \citenamefont {Gullans}, \citenamefont {Gorshkov}, \citenamefont
		{Thompson}, \citenamefont {Chin}, \citenamefont {Lukin},\ and\ \citenamefont
		{Vuleti{\'c}}}]{Liang2018}%
	\BibitemOpen
	\bibfield  {author} {\bibinfo {author} {\bibfnamefont {Q.-Y.}\ \bibnamefont
			{Liang}}, \bibinfo {author} {\bibfnamefont {A.~V.}\ \bibnamefont
			{Venkatramani}}, \bibinfo {author} {\bibfnamefont {S.~H.}\ \bibnamefont
			{Cantu}}, \bibinfo {author} {\bibfnamefont {T.~L.}\ \bibnamefont
			{Nicholson}}, \bibinfo {author} {\bibfnamefont {M.~J.}\ \bibnamefont
			{Gullans}}, \bibinfo {author} {\bibfnamefont {A.~V.}\ \bibnamefont
			{Gorshkov}}, \bibinfo {author} {\bibfnamefont {J.~D.}\ \bibnamefont
			{Thompson}}, \bibinfo {author} {\bibfnamefont {C.}~\bibnamefont {Chin}},
		\bibinfo {author} {\bibfnamefont {M.~D.}\ \bibnamefont {Lukin}}, \ and\
		\bibinfo {author} {\bibfnamefont {V.}~\bibnamefont {Vuleti{\'c}}},\
	}\bibfield  {title} {\enquote {\bibinfo {title} {Observation of three-photon
				bound states in a quantum nonlinear medium},}\ }\href {\doibase
		10.1126/science.aao7293} {\bibfield  {journal} {\bibinfo  {journal}
			{Science}\ }\textbf {\bibinfo {volume} {359}},\ \bibinfo {pages} {783--786}
		(\bibinfo {year} {2018})}\BibitemShut {NoStop}%
\end{thebibliography}

\begin{thebibliography}{2}%
	\makeatletter
	\providecommand \@ifxundefined [1]{%
		\@ifx{#1\undefined}
	}%
	\providecommand \@ifnum [1]{%
		\ifnum #1\expandafter \@firstoftwo
		\else \expandafter \@secondoftwo
		\fi
	}%
	\providecommand \@ifx [1]{%
		\ifx #1\expandafter \@firstoftwo
		\else \expandafter \@secondoftwo
		\fi
	}%
	\providecommand \natexlab [1]{#1}%
	\providecommand \enquote  [1]{``#1''}%
	\providecommand \bibnamefont  [1]{#1}%
	\providecommand \bibfnamefont [1]{#1}%
	\providecommand \citenamefont [1]{#1}%
	\providecommand \href@noop [0]{\@secondoftwo}%
	\providecommand \href [0]{\begingroup \@sanitize@url \@href}%
	\providecommand \@href[1]{\@@startlink{#1}\@@href}%
	\providecommand \@@href[1]{\endgroup#1\@@endlink}%
	\providecommand \@sanitize@url [0]{\catcode `\\12\catcode `\$12\catcode
		`\&12\catcode `\#12\catcode `\^12\catcode `\_12\catcode `\%12\relax}%
	\providecommand \@@startlink[1]{}%
	\providecommand \@@endlink[0]{}%
	\providecommand \url  [0]{\begingroup\@sanitize@url \@url }%
	\providecommand \@url [1]{\endgroup\@href {#1}{\urlprefix }}%
	\providecommand \urlprefix  [0]{URL }%
	\providecommand \Eprint [0]{\href }%
	\providecommand \doibase [0]{http://dx.doi.org/}%
	\providecommand \selectlanguage [0]{\@gobble}%
	\providecommand \bibinfo  [0]{\@secondoftwo}%
	\providecommand \bibfield  [0]{\@secondoftwo}%
	\providecommand \translation [1]{[#1]}%
	\providecommand \BibitemOpen [0]{}%
	\providecommand \bibitemStop [0]{}%
	\providecommand \bibitemNoStop [0]{.\EOS\space}%
	\providecommand \EOS [0]{\spacefactor3000\relax}%
	\providecommand \BibitemShut  [1]{\csname bibitem#1\endcsname}%
	\let\auto@bib@innerbib\@empty
	%</preamble>
	\bibitem [{\citenamefont {Kashcheyevs}\ and\ \citenamefont
		{Kaestner}(2010)}]{Kashcheyevs2016S}%
	\BibitemOpen
	\bibfield  {author} {\bibinfo {author} {\bibfnamefont {V.}~\bibnamefont
			{Kashcheyevs}}\ and\ \bibinfo {author} {\bibfnamefont {B.}~\bibnamefont
			{Kaestner}},\ }\bibfield  {title} {\enquote {\bibinfo {title} {Universal
				decay cascade model for dynamic quantum dot initialization},}\ }\href
	{\doibase 10.1103/PhysRevLett.104.186805} {\bibfield  {journal} {\bibinfo
			{journal} {Phys. Rev. Lett.}\ }\textbf {\bibinfo {volume} {104}},\ \bibinfo
		{pages} {186805} (\bibinfo {year} {2010})}\BibitemShut {NoStop}%
	\bibitem [{\citenamefont {Muthukrishnan}\ \emph {et~al.}(2004)\citenamefont
		{Muthukrishnan}, \citenamefont {Agarwal},\ and\ \citenamefont
		{Scully}}]{Muthukrishnan2004S}%
	\BibitemOpen
	\bibfield  {author} {\bibinfo {author} {\bibfnamefont {A.}~\bibnamefont
			{Muthukrishnan}}, \bibinfo {author} {\bibfnamefont {G.~S.}\ \bibnamefont
			{Agarwal}}, \ and\ \bibinfo {author} {\bibfnamefont {M.~O.}\ \bibnamefont
			{Scully}},\ }\bibfield  {title} {\enquote {\bibinfo {title} {Inducing
				disallowed two-atom transitions with temporally entangled photons},}\ }\href
	{\doibase 10.1103/PhysRevLett.93.093002} {\bibfield  {journal} {\bibinfo
			{journal} {Phys. Rev. Lett.}\ }\textbf {\bibinfo {volume} {93}},\ \bibinfo
		{pages} {093002} (\bibinfo {year} {2004})}\BibitemShut {NoStop}%
	\bibitem [{\citenamefont {Albrecht}\ \emph {et~al.}(2019)\citenamefont
		{Albrecht}, \citenamefont {Henriet}, \citenamefont {Asenjo-Garcia},
		\citenamefont {Dieterle}, \citenamefont {Painter},\ and\ \citenamefont
		{Chang}}]{Albrecht2019S}%
	\BibitemOpen
	\bibfield  {author} {\bibinfo {author} {\bibfnamefont {A.}~\bibnamefont
			{Albrecht}}, \bibinfo {author} {\bibfnamefont {L.}~\bibnamefont {Henriet}},
		\bibinfo {author} {\bibfnamefont {A.}~\bibnamefont {Asenjo-Garcia}}, \bibinfo
		{author} {\bibfnamefont {P.~B.}\ \bibnamefont {Dieterle}}, \bibinfo {author}
		{\bibfnamefont {O.}~\bibnamefont {Painter}}, \ and\ \bibinfo {author}
		{\bibfnamefont {D.~E.}\ \bibnamefont {Chang}},\ }\bibfield  {title} {\enquote
		{\bibinfo {title} {Subradiant states of quantum bits coupled to a
				one-dimensional waveguide},}\ }\href {\doibase 10.1088/1367-2630/ab0134}
	{\bibfield  {journal} {\bibinfo  {journal} {New J. Phys.}\ }\textbf {\bibinfo
			{volume} {21}},\ \bibinfo {pages} {025003} (\bibinfo {year}
		{2019})}\BibitemShut {NoStop}%
	\bibitem [{\citenamefont {Zhang}\ and\ \citenamefont
		{M\o{}lmer}(2019)}]{Molmer2019S}%
	\BibitemOpen
	\bibfield  {author} {\bibinfo {author} {\bibfnamefont {Y.-X.}\ \bibnamefont
			{Zhang}}\ and\ \bibinfo {author} {\bibfnamefont {K.}~\bibnamefont
			{M\o{}lmer}},\ }\bibfield  {title} {\enquote {\bibinfo {title} {Theory of
				subradiant states of a one-dimensional two-level atom chain},}\ }\href
	{\doibase 10.1103/PhysRevLett.122.203605} {\bibfield  {journal} {\bibinfo
			{journal} {Phys. Rev. Lett.}\ }\textbf {\bibinfo {volume} {122}},\ \bibinfo
		{pages} {203605} (\bibinfo {year} {2019})}\BibitemShut {NoStop}%
	\bibitem [{\citenamefont {Zhang}\ \emph {et~al.}(2019)\citenamefont {Zhang},
		\citenamefont {Yu},\ and\ \citenamefont {M{\o}lmer}}]{zhang2019subradiantS}%
	\BibitemOpen
	\bibfield  {author} {\bibinfo {author} {\bibfnamefont {Y.-X.}\ \bibnamefont
			{Zhang}}, \bibinfo {author} {\bibfnamefont {C.}~\bibnamefont {Yu}}, \ and\
		\bibinfo {author} {\bibfnamefont {K.}~\bibnamefont {M{\o}lmer}},\ }\bibfield
	{title} {\enquote {\bibinfo {title} {Subradiant dimer excited states of atom
				chains coupled to a {1D} waveguide},}\ }\href
	{https://arxiv.org/abs/1908.01818} {\bibfield  {journal} {\bibinfo  {journal}
			{arXiv:1908.01818}\ } (\bibinfo {year} {2019})}\BibitemShut {NoStop}%
	\bibitem [{\citenamefont {Poshakinskiy}\ and\ \citenamefont
		{Poddubny}(2016)}]{Poshakinskiy2016S}%
	\BibitemOpen
	\bibfield  {author} {\bibinfo {author} {\bibfnamefont {A.~V.}\ \bibnamefont
			{Poshakinskiy}}\ and\ \bibinfo {author} {\bibfnamefont {A.~N.}\ \bibnamefont
			{Poddubny}},\ }\bibfield  {title} {\enquote {\bibinfo {title}
			{Biexciton-mediated superradiant photon blockade},}\ }\href {\doibase
		10.1103/PhysRevA.93.033856} {\bibfield  {journal} {\bibinfo  {journal} {Phys.
				Rev. A}\ }\textbf {\bibinfo {volume} {93}},\ \bibinfo {pages} {033856}
		(\bibinfo {year} {2016})}\BibitemShut {NoStop}%
\end{thebibliography}

\end{document}